\documentclass[a4paper,11pt]{article}
\usepackage{arxiv}
\usepackage[utf8]{inputenc} 
\usepackage[numbers]{natbib}

\usepackage{authblk}
\usepackage{orcidlink}
\usepackage{amsmath}
\usepackage{float}

\newcommand{\stateful}{\emph{Stateful}}
\newcommand{\STACK}{STACK}
\newcommand{\SDE}{SDE}
\newcommand{\clusterA}{$\mathcal{A}$}
\newcommand{\clusterB}{$\mathcal{B}$}
\newcommand{\clusterC}{$\mathcal{C}$}
\newcommand{\varEmph}[1]{\texttt{#1}}

\newcommand{\nicefrac}[2]{#1 / #2\,}

\begin{document}

\title{Student behaviour and engagement with adaptive exercises on a thermodynamics course}

\author[1]{Matti Harjula\,\orcidlink{0000-0002-0622-5177}}
\author[2]{Ville Havu}
\author[3,4]{Inkeri Kontro\,\orcidlink{0000-0002-1874-3756}}
\author[4]{Kimmo Kulmala\,\orcidlink{0000-0003-2706-0022}}
\author[1]{Jarmo Malinen\,\orcidlink{0000-0001-9117-8450}}
\author[2]{Petri Salo\,\orcidlink{0000-0002-7749-9116}}

\affil[1]{Department of Mathematics and Systems Analysis, Aalto University}
\affil[2]{Department of Applied Physics, Aalto University}
\affil[3]{Physics Unit, Tampere University}
\affil[4]{Department of Physics, University of Helsinki}

\begin{abstract}
A teaching experiment was carried out in a university-level thermodynamics course using adaptive and interactive e-learning material, created in the new Moodle question type Stateful extending the original e-learning platform STACK. The system collects data about the students that is used to algorithmically classify them according to their behaviour in solving problems. It is observed that the classification of this data predicts students' success in the other parts of the course for a majority of students. Also, the classification is statistically consistent with Thermodynamic Concept Survey and Maryland Physics Expectation Survey.
\end{abstract}

\maketitle

\section{\label{IntroSec} Introduction}

This article discusses the suitability of a novel e-learning concept for teaching
university-level elementary physics. The motivation and some technical details of the
pedagogical experiment are that the use of any e-learning
platform has a number of potential benefits. Such benefits incluse better reaching
the students and giving them support without the need of assistants
for scheduled learning, thus providing the students with interactive
study time outside usual office hours.  The adaptivity of the study
material to different students' needs is an important issue as
well. It should, however, be noted that some of the students do not
want to interact with a computerised system for reasons that are
poorly understood. Realising these benefits and taking into account
the shortcomings requires continuous development of technological
platforms and study materials built on them. It is crucial to study
the platform in an authentic setting, and we expect the novel, more
adaptive exercises to extend individualised feedback from its current
scope in e-learning, rather than to replace the interaction between
the teacher and the student.

The novel e-learning exercise in this work is based on \stateful{} which is a new extension of the
traditional e-learning platform \STACK{} (System for Teaching and Assessment using
a Computer algebra Kernel).  \STACK{} is a popular
e-learning environment for interactive materials that is suitable for
teaching general mathematical content, and it provides feedback for
students and automatic assessment of students' answers for
teachers. Numerous applications and a wide literature exists about
\STACK{}; see, e.g.,
\cite{2013CAA,R-M-T:OAACU,2017STACKUnits,paiva2017intelligent,kulmala2021,erskine_mestel_2018,Nakamura2013,Nakamura2014}.

Individually randomised and interactive exercises on the \STACK{}
platform have been provided for university students for many years
\cite{2013CAA}. While the \STACK{} platform can offer randomisation of
initial values and individualised feedback based on the students'
answers, all students, however, see the same problem and are expected
to solve it following the essentially same steps.  \stateful{} was
originally drafted as a question type of \STACK{} in
\cite{harjula2016stack}, and it makes it possible to interactively
link traditional \STACK{} exercises as separate \emph{scenes} in a
network; see also \cite{altieri_mike_2020_3944786}.  More precisely, the transition to the next scene is
determined by rules in the decision tree according to which students'
answers in earlier scenes are classified\footnote{\stateful{} is a
  technology reminiscent of the decades old ``rogue-like'' game
  engines for text-based adventure games such as Zork. The \stateful{}
  Diesel exercise considered here, however, is far too restricted to
  be regarded as a true e-learning game.}.  The \stateful{} question
type makes it feasible to create e-learning materials that have an
underlying story or a plot, emphasising strategic skills in problem
solving while adapting to varying levels of students'
skills. Students' answers change the internal state of the exercise
and take the students to different paths. This is how \stateful{}
exercises implement game-like characteristics in non-game settings,
thus making a gamified learning experience attainable
\cite{deterding2011game}. It is worth pointing out that this mode of
learning is in stark contrast to drilling for procedural fluency or
checking factual knowledge by, e.g., Multiple Choice Questions (MCQ)
as discussed in \cite{R-M-T:OAACU}.

Gamified learning experiences have been used to foster engagement to
physics for primary school students and in science fairs
\cite{vieyra2020gamified}. Even though the idea of using game-like
interactivity for teaching mathematical subjects is by no means new
(see, e.g., \cite{devlin2011mathematics,SUN2021100457}), little is
known about how such pedagogical methods are received by university
students. However, many university students may find game-like reward
systems motivating as well \cite{rincon2021gamification,Dichev:2017,Immonen23}.

The e-learning material of this study is a single \stateful{} Diesel
Exercise (henceforth, \SDE{}) that describes thermodynamics of a
four-stroke Diesel engine, omitting the actual mechanical
function. Such an engine is particularly suitable as a subject for
\stateful{} question type involving multiple scenes. Not only does the
action of the engine consist of four strokes but also its
thermodynamical state diagram comprises four stages where
understanding of different thermodynamical processes is required. For
the descriptions of the Diesel cycle and the corresponding \SDE{},
see~Sections~\ref{DieselSec}~and~\ref{ExerciseSec} below.

The architecture of the \stateful{} question type makes it possible to
analyse the learning process based on the collected data. As shown by
the teaching experiment reported in this article, students interact
with \stateful{} in a number of different ways but far from
randomly. \stateful{} produces large amounts of behavioural data about
students in their search for the correct answer and deeper
understanding of the subject matter. The analysis of this data appears
to be a great challenge both pedagogically and technically. Modern
data science tools are helpful for finding the signal in noise
\cite{Romero2008DataMA,Romero2013,romero2020educational,Algarni2016DataMI}.
The proposed approach takes into account the necessary components of
learning analytics: data, its analysis, and a mechanism for action to
improve the educational environment; see, e.g.,
\cite{Ebner2018,Ebner19:LearningAnalytics,Wright2014}.

The underlying pedagogical challenge in teaching thermodynamics is to 
identify problems in studies and engage
all beginning students in learning university physics, similarly
to \cite{Wright2014,WANG202242,Huberth2015}. We
are interested in both learning outcomes and attitudes of the students
while keeping the focus in \SDE{}. For this purpose, a classification
of students' \emph{behavioural patterns} is carried out by
Self-Organizing Maps (SOM) \cite{KOHONEN201352}, based only on their
logged actions in solving \SDE{}. As outcome of this process, three student groups \clusterA{}, \clusterB{}, and \clusterC{} are discovered.  For different classification and
clustering algorithms used for Educational Data Mining (EDM), see,
e.g., \cite{Romero2013,krizanic:2020}. Independently of the clustering
approach, a parametric model (i.e., \emph{fitted metric}) is proposed in this article
so as to optimally predict students' success in other parts of the
course, based only on the data from \SDE{}; see also
\cite{math10203758,mci/Cocea2006}. In this article, we approach the
general research problem from three research questions:
\begin{enumerate}
\item \label{FirstIssues} To what extent do the emerging three major
  student groups \clusterA{}, \clusterB{}, and \clusterC{} statistically differ from each other? How does the
  membership in these groups predict the success in other parts of the
  course, including the final grade?  In which ways do the students in
  different groups benefit (or not benefit) from \SDE{}?
\item \label{SecondIssues} Does the fitted metric model have
  predictive power on the success of students in the course?  Does the
  fitted metric support the robustness of classification produced by
  SOM?
\item \label{ThirdIssues} Based on profiling by Thermodynamic Concept
  Survey (TCS) and Maryland Physics Expectation Survey (MPEX), how do
  the proficiencies and attitudes statistically relate with groups
  \clusterA{}, \clusterB{}, and \clusterC{}?  Is the development in
  students' TCS profiles during the course related to which of these
  groups they belong?
\end{enumerate}
It is observed in this work that the data from \stateful{} provides a
joint picture of student's behaviour and learning outcome which is
quite consistent when studied from all three directions
above. In particular, the logged data is rich enough for analysis by
fitted metric modelling for groups \clusterA{} and \clusterB{} but not
for students in group \clusterC{}, who typically do not complete
\SDE{}. Details and statistics are given in Section~\ref{ResultsSec}.

The outline of the article is as follows: In
Section~\ref{BackgroundSec}, we describe the background of the study:
the pedagogical background, the university course, and the instruments
used. Section~\ref{StatefulSec} introduces the \stateful{} question
type as well as scoring and design principles of exercise
material. Section~\ref{MaterialSec} presents the test subjects and the
data collected, Section~\ref{MethodsSec} the data analysis, and
Section~\ref{ResultsSec} the results and discussion. Conclusions are
given in Section~\ref{DiscConcSec}.

\section{\label{BackgroundSec} Background}

\subsection{\label{PedaSec} Pedagogical background}
    
Demands for more engaging education in science, technology,
engineering, and mathematics (i.e., the STEM subjects) have been
increasingly sounded in recent years. The call is twofold: the society needs more
STEM students for the increasingly technological society, which means the number of students increases. Both due to this and as a goal in its own right, the engineering
profession attracts more diverse students than before. The
engineering education should be accordingly reformed to improve
learning outcomes and engagement \cite{Joyce2011}. Quality instruction
is required to achieve these goals.

Instructional strategies address the need for quality in the physical
or virtual classroom. In particular, \emph{active learning} comprises
instruction strategies that focus on meaningful student activities and
involve student reflection \cite{Prince2004}. These strategies lead to
higher learning outcomes than traditional methods
\cite{Prince2004,Freeman2014}. Similarly, research consistently shows
that \emph{cooperative learning} increases students' efforts,
learning, and it improves student well-being \cite{Prince2004,
  Johnson2014}.

Another crucial factor in improving learning outcomes is feedback and
interaction \cite{Hattie2007,Shute2008}. \emph{Formative feedback}
guides the student forward, and it can also give the student the
opportunity to revise their previous work \cite{Shute2008}. The
development of computerised problem-solving tasks and assessment
provide an opportunity to give students feedback tailored according to
their individual answers. \emph{Elaborative feedback} gives
explanations, more material or hints \cite{Shute2008}, and improves
learning outcomes more than simply informing the student on the
correctness of the answer \cite{Attali2015}. Traditionally, much of
the elaborative feedback has come from the teacher through markings on
students' solutions on paper. From a purely pragmatic point of view,
computerised feedback and assessment frees instructor resources that
can be reallocated to, e.g., contact teaching and materials
development.

\subsection{\STACK{}}

\STACK{} is an automatic assessment system capable for advanced feature
detection of students' answers. By feature detection we understand the
ability to test for various properties of the answer data, starting
from the numerical equivalence within a given tolerance and the
elementary pattern matching of strings. Detecting the algebraic
equivalences such as the equalities in the expression
\begin{equation*}
  x^3 - 6x^2 + 11x - 6 = \left(x - 3\right)\left(x^2 - 3x + 2\right) 
  = \left(x - 3\right)\left(x - 2\right)\left(x - 1\right)
\end{equation*}
is quite a demanding requirement for any assessment system. In \STACK{},
these manipulations are realised by making use of the Computer Algebra
System (CAS) Maxima \cite{maxima} that has been integrated into the
platform. \STACK{} provides tools for processing significant figures
rules and answer data having physical dimensions
\cite{2017STACKUnits}. Indeed, \STACK{} understands \(6.00 \textrm{in}\)
to be equivalent with \(0.500 \textrm{ft}\) and that these quantities
are given above with the comparable accuracy in terms of significant
figures.

\STACK{} questions are typically parametric, and the parameter values
can be randomly selected from a predefined set. Hence, different
students may receive different variants of the same question. The
parameter values can be introduced to the question text, and they are
used for assessing students' answer data. Model answers can be
generated for each student separately, depending on the specifically
parameterised version the student has tried to answer.

Many types of input fields are supported in \STACK{}, ranging from
numerical values to matrices and complex algebraic expressions.
Multiple input fields, interleaved by question text, are supported.
Filling in a subset of the inputs may trigger the processing multiple
Potential Response Trees (PRT) which is the aggregate of conditional
statements that define the answer data processing logic of a \STACK{}
question.  More precisely, the PRT is a binary decision tree for
feature detection where each node executes a simple test on the answer
data and the question parameters.  Based on the result, the specific
feedback can be given to the student, and the grading is adjusted in
\STACK{}.

Automatic assessment systems have been successfully used in
conjunction with more traditional teaching methods in a way that
improves each of the four dimensions of e-learning described in
Section~\ref{PedaSec} (i.e., active and cooperative learning,
formative and elaborative feedback). The novel \stateful{} question
type (discussed in Section~\ref{StatefulSec} below) provides a
suitable tool to give students automated, personalised, and versatile
feedback beyond the capabilities of the traditional \STACK{}. Many
automatic assessment systems have the potential of producing
overwhelming amounts of raw student answer data well beyond the
requirements of merely grading the coursework. Our purpose is to study
to what extent \stateful{} produces useful data for \emph{learning
  analytics}, which can be seen as an additional dimension of
e-learning as well.

\subsection{\label{TeachExpSec} Course arrangements and  grading}

The e-learning experiment was arranged in conjunction with an
undergraduate five-week physics course for first-year students
majoring in mechanical engineering in a Finnish university. Fluid
dynamics and theory of elasticity were taught on one week each. The
remaining three weeks dealt with traditional thermodynamics in the
following subjects: kinetic theory of gases, thermodynamic processes,
heat engines, first and second law of thermodynamics, and heat
conduction.  The textbook \emph{Physics for Scientists \& Engineers
  with Modern Physics} by Douglas Giancoli
\cite{GiancoliDouglasC2008Pfsa} was followed.

The teaching was arranged as ten lectures (two times 45 minutes each,
held online due to COVID-19 pandemic in 2020) and equally many
exercise sessions of the same length. All exercises were graded
automatically in classical \STACK{} except for one session that was
reserved for the e-learning experiment. This experiment made use of a
single, extensive exercise implemented with \stateful{}. The exercise
is referred to as \SDE{} in this article, and it involves a number of
aspects of the Diesel cycle reviewed below in
Section~\ref{DieselSec}. Moreover, two laboratory sessions were
organised: one in the student laboratory and the other online (again
due to COVID-19 pandemic).  In addition to the exercise sessions, the
students had pre-lecture exercises for self-study implemented in the
\emph{MasteringPhysics} \cite{masteringphysics} platform and weekly learning diaries as
post-lecture assignments, which were graded by the instructor.

Considering the final grade of the course, the exam and exercises
accounted for 30\% each, the laboratory sessions for 20\%, and the
pre- and post-lecture work accounted for 10\% each.  For the analysis
carried out in this article, only the exercises and the exam are used
since they produce fine-grained numerical information on student
behaviour that is directly suitable for studies by quantitative
methods. The remaining course activity, representing 40\% of the final
grade, are less aligned with the learning goals of \SDE{}.

In autumn semester 2020, 182 students enrolled to the course, 163
participated in the final exam, and 151 passed. Due to
COVID-19 pandemic, the exam was held online in ``open-book'' format
where the students had access to all the material. However, the
results of the exam were very similar to results from previous years
when traditional exams were arranged. Due to the online exam, the course
results were exceptionally entered into study registry using the
pass/fail -scale even though the numerical grading was internally
carried out as usual.

\subsection{\label{ExExSec} Exercises and exam}

\begin{figure}
  \centering
    \includegraphics[width=0.65\textwidth]{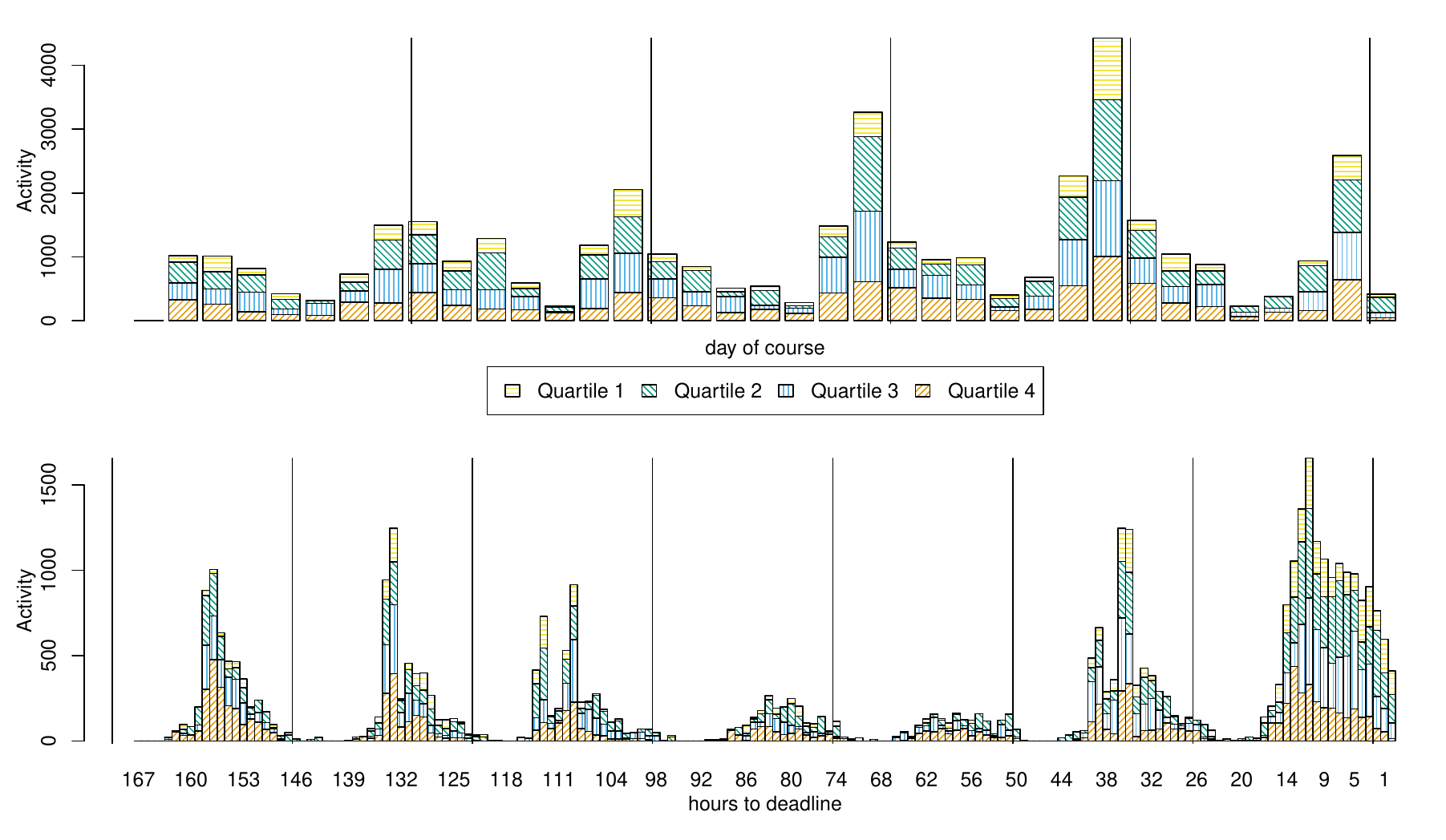}\includegraphics[width=0.35\textwidth,height=6.5cm]{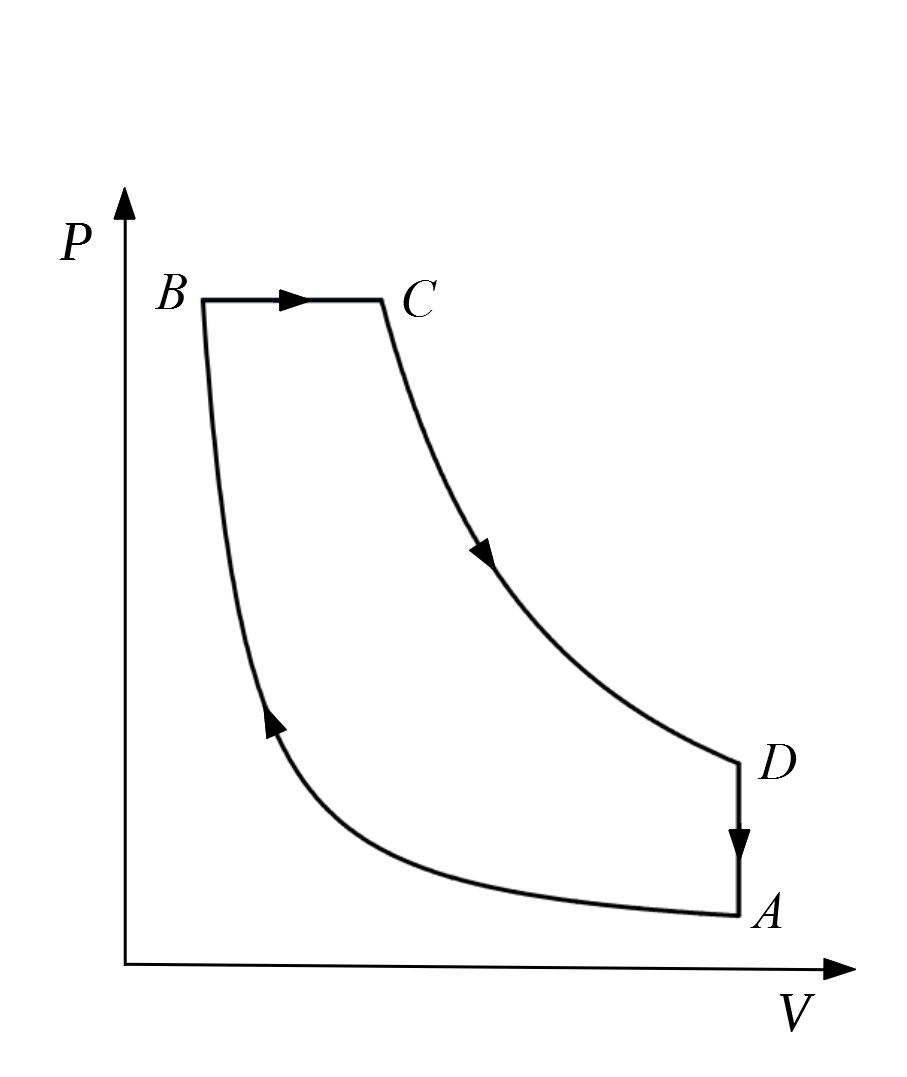}
    \caption{Left panels: Student activity of \STACK{} exercises and
      \SDE{} measured by logged submissions.  The time axis is given
      in relation to deadlines in two different ways. In the top
      panel, daily submissions are shown during the five weeks where the vertical lines mark the weekly deadlines over the
      duration of the course.  In the bottom panel, the time axis
      gives hours before the deadline over the interval of one week, aggregated over all five weeks. The vertical lines denote the midnight. The lower activity
      on weekends is visible, and the large spikes correlate with
      guided exercise sessions. In both panels, the student population
      has been divided into quartiles based on their success in the
      exam, quartile 4 consisting of most successful.  Right panel:
      Diesel cycle as a $PV$-diagram.}
    \label{Diesel-cycle}
\end{figure}

Each of the weekly exercise sessions comprised six traditional
\STACK{} questions with the exception of the 4th course week when
\SDE{} and three \STACK{} questions were used. Each \STACK{} question
had the value of 1\% of the course total points whereas the more
demanding Diesel cycle was worth 3\%. The students were allowed an
unlimited number of tries for all exercises, and no deductions were
applied for trying to solve the same exercise several times. The
weekly deadline for returning the solutions was set to Wednesday morning at 3~a.m.; see
Fig.~\ref{Diesel-cycle} (left panel) for student activity as a
function of time.

In contrast to the \STACK{}/\stateful{} exercises, the final exam did
not involve solving computational problems but focused on
thermodynamical concepts and on students' ability to apply their
skills to analyse thermodynamical phenomena. However, use of equations
to explain the concepts was required.  There were six problems in the
exam, four of which had a connection to traditional
thermodynamics. There was one problem in the exam that was closely
related to the Diesel cycle, where the students were
given a set of thermodynamic processes and asked to use them to
construct a thermodynamic cycle for a heat engine. However,
reconstructing the Diesel cycle was not a valid answer since the set
did not contain the required processes.  The student was also required
to analyse the exchange of heat and work in each of the sub-processes
of the engine they proposed. The three other problems of thermodynamics
dealt with adiabatic processes and the First Law of Thermodynamics,
the Second Law of Thermodynamics and a perpetual motion machine, and
heat conduction in a two-layer insulating structure. The total points
of the exam are described by the statistical variable
\varEmph{ExamTotal} in this article.

\subsection{\label{InstSec} Description of instruments}

In this article, we study the reactions of students and learning
outcomes associated with the \stateful{} question type. For this
purpose, two kinds of quantitative instruments are required. Firstly,
validated instruments (such as MPEX and TCS below) are used for
evaluating students' attitudes and propensities. Secondly, evaluation
and classification tools are needed for quantifying the success in
problem solving and the details of the problem solving process, based
on the multimodal data from the course systems.

Conceptual learning is an important learning outcome which is separate
from procedural knowledge \cite{brown2009conceptual}. To measure the
conceptual knowledge, we administered the Thermodynamic Concept Survey
(TCS) \cite{WattanakasiwichTCS}, which has previously been translated
to Finnish and used in university-level thermophysics courses \cite{koskinen2018primetime}.  We carried out the TCS
survey twice, namely on weeks three and five of the course (i.e.,
before and after \SDE{}) to measure the base level and changes in
students' conceptual competences.  Questions 1--7 from the TCS survey
were identified as \varEmph{PriorKnowledge}, as these topics were not
part of the course.  Questions 8--35 covered the course material
itself, and we denote their total sum as the statistical variable
\varEmph{Relevant}. Based on content, there are also two subgroups of questions that are
important when considering the Diesel cycle, giving similarly rise to
statistical variables \varEmph{Adiabatic} (consisting of TCS questions
13-19) and \varEmph{Cyclic} (consisting of TCS questions 20-31)
describing processes in Thermodynamics. There are two copies of each
of the four statistical variables from TCS, namely
\varEmph{Adiabatic}, \varEmph{Cyclic}, \varEmph{PriorKnowledge}\footnote{Technically, the variable \varEmph{post-PriorKnowledge} exists but is ignored.}, and
\varEmph{Relevant}, referring to the TCS carried out before and after
\SDE{}.  These variables are denoted by, e.g., \varEmph{pre-Adiabatic}
and \varEmph{post-Adiabatic}, and so on.

Student beliefs are another important factor in learning and, in
particular, in successful problem solving
\cite{Ramirez-Arellano2019,Adams2015}. Attitudes can concern specific
topics, but also the general view on learning is important. Attitudes
towards physics are often characterised on a scale from
non-expert-like to expert-like and surveyed by instruments such as
Maryland Physics Expectation Survey (MPEX) \cite{RedishMPEX} or
Colorado Learning Attitudes about Science Survey (CLASS)
\cite{adams2006}. Pre-instruction student attitudes have a unique,
though moderate effect on learning both when measured by conceptual
gains or by exam averages \cite{kortemeyer2007,ding2014,cahill2018}.
We used the MPEX survey to study the effect of student attitudes on
the engagement in instruction and learning.  MPEX consists of 34
statements which are answered on a 5-point Likert scale (strongly
disagree to strongly agree). The MPEX was scored in the standard way
of counting the number of answers aligning with expert opinion
\cite{RedishMPEX}, and the corresponding statistical variable is
\varEmph{MPEXTotal} in this article. The MPEX survey was translated to Finnish
twice by two different translators, after which the translations were compared and
the disagreements were resolved by discussion. The resulting translation 
was translated back into English by a third person, and the result was compared
with the original survey wordings. No inconsistencies were found in translation.

Let us next consider the evaluation of learning outcome and process.
Exams are the traditional way of quantifying learning outcome in
university-level education, but continuous assessment has been
increasingly used during the last few decades. The emergence of
e-learning systems (such as \STACK{} and \stateful{}) has facilitated
continuous assessment but also revealed new challenges how to arrange
such assessment in an effective and fair manner.  Classical \STACK{}
questions can be assessed in term of ``raw points'' (as defined by the
question logic) and the number of attempts the student makes.  In
addition to this, there is \emph{path information} available in
\stateful{}, describing the order the student has visited or skipped
scenes of \SDE{}. The path information can be numerically quantified
and combined with other types of performance data by statistical means
and optimisation.  As a result of this approach, we propose a new
instrument which we call \emph{fitted metric}.  We point out that the
purpose of the fitted metric is not to grade students' performance in
an obscure manner but, instead, to produce deeper understanding and
richer learning analytics data in conjunction with other evaluation
and psychometric tools. Moreover, the fitted metric can be used to
evaluate the usefulness of the data produced by a \stateful{}
exercise.

The statistical computing was done using the R-system v.4.1 which was
extended by \texttt{kohonen-package} v.3.0 for including the SOM
algorithms that were used with default settings and a rectangular
adjacency grid. \stateful{} v.1.0.2 and \STACK{} v.4.3.5 were used on
a Moodle environment based on v.3.8.2+.  The optimal parameter
estimation for fitted metric modelling was carried out using Octave
v.6.3.0. The e-learning material, i.e., \SDE{}, was created in the
experimental development environment Eleaga Editor v.r1906.

\subsection{\label{DieselSec} Diesel cycle}

The e-learning experiment involved a single \stateful{} exercise about
the Diesel cycle which serves as an exemplar of a non-trivial
thermodynamic cycle for the purposes of the course. A Diesel cycle is
a thermodynamical cycle used to model a four-stroke Diesel
engine. However, the mechanical operating principle of a concrete
Diesel engine was not emphasised in the course.  We complete this
section by giving a reminder of the Diesel cycle to the reader.

The pressure-volume-diagram ($PV$-diagram) of the Diesel cycle is
shown in Fig.~\ref{Diesel-cycle} (right panel). It consists of
adiabatic compression, isobaric expansion, adiabatic expansion, and
isochoric cooling. In the first adiabatic process, the fresh air in
the cylinder is compressed adiabatically by a movement of the piston
(${A} \rightarrow {B}$). The following isobaric
expansion corresponds to burning of the fuel injected into the
cylinder (${B} \rightarrow {C}$) where heat is added to
the system. After all injected fuel has been consumed, the piston
moves in a power strike modelled by an adiabatic expansion
(${C} \rightarrow {D}$). In the last phase, the piston
moves once between its extrema, expelling the exhaust gases from the
cylinder and taking in fresh air. This is modelled by an isochoric
cooling, and the systems gives out heat in this process (${D}
\rightarrow {A}$).

The changes of pressure and volume of the gas during the adiabatic
process are governed by $P_{A} V_{A}^\gamma =
P_{B} V_{B}^\gamma$ and $P_{C}
V_{C}^\gamma = P_{D} V_{D}^\gamma$. Since air is
approximated by the diatomic ideal gas the adiabatic constant is
$\gamma = 1.4$.  The pressure is constant $P_{B} =
P_{C}$ in the isobaric process, and the volume of the system is
constant $V_{D} = V_{A}$ during the isochoric
process. It is further assumed that the gas in the cylinder follows
the ideal gas law and satisfies the equation of state $PV = nRT$ for
every state of the system during the Diesel cycle. Here $n$ is the
amount of the substance in the system, $T$ is the absolute temperature
of the system, and $R$ is the molar gas constant. The heat added to
the system during the isobaric expansion can be obtained from $Q_H =
nC_P \left( T_{C} - T_{B} \right)$, where $C_P =
\nicefrac{7}{2}R$ is the isobaric molar heat capacity of the diatomic
ideal gas. Similarly the heat expelled during the isochoric cooling
can be obtained from $Q_L = nC_V \left( T_{D}-T_{A}
\right)$, where $C_V = \nicefrac{5}{2}R$ is the isochoric molar heat
capacity of the diatomic ideal gas. The thermodynamic efficiency of
the cycle is $\epsilon = \nicefrac{W}{Q_H}$, where $W = Q_H - Q_L$ is the
work done by the cycle. This work is equal to the area enclosed by the
cycle in the $PV$-diagram.

In a typical problem setting the student is given the amount of
substance in the system and temperature and pressure at one point of
the cycle (most often at the point ${A}$ corresponding to the
ambient conditions). Furthermore the compression ratio $r_c =
\nicefrac{V_{A}}{V_{B}}$ and the expansion ratio $r_e =
\nicefrac{V_{D}}{V_{C}}$ are provided to describe the limits
of the movement of the piston in the cylinder. The student is first
asked to calculate pressure and temperature at each of the corner
points of the cycle (${B},{C},{D}$) as well as
the heat added to and expelled from the system during one
cycle. Finally, the student is asked to compute the work done and the
thermodynamical efficiency of the Diesel cycle.

\section{\label{StatefulSec}  \stateful{}}

\stateful{} is a new question type for Moodle that uses \STACK{}
\cite{2013CAA} as its engine and for defining the question
semantics. The primary distinctive feature of \stateful{} is the
following: The student's action may affect the internal state of the
question, thus leading to interactive adaptations seen by the student.
Since the question type of traditional \STACK{} does not have an
internal state, the similar adaptation -- resembling a form of an
intelligent dialogue at its best -- is not possible. Because
\stateful{} is experimental technology at the writing of this article,
we briefly review the basic concepts below.

The \stateful{} question type is based on the exploratory work in
\cite{harjula2016stack} where \STACK{} was crudely augmented with state
variables, allowing questions to change shape based on student's
interaction. This idea appeared to be attractive and feasible but the
authoring of exercises proved to be overly complicated. For this
reason, custom tools for authoring and testing were developed, and the
question type was streamlined as well.

A \stateful{} exercise consists of multiple parametric \STACK{}-like
questions called \emph{scenes}. Compared to the traditional \STACK{},
the PRTs of scenes have been somewhat enriched.  One of the scenes is
the entry point of the exercise, and exactly one scene is being
presented to the student at any moment. The parameter values in scenes
can be randomised, or they can be tied to state variables.  At the end
of a PRT branch of a scene, a \emph{state transition} may be triggered
which leads to a change in parameter values or to a transition to
another scene. We point out that the values of parameters and state
variables store information of student's past inputs.

From the computer science point of view, \stateful{} exercises are
state machines that accept external control.  While the \stateful{}
question type is not a framework for producing video games with fancy
animations \cite{devlin2011mathematics}, it does allow the student to
observe the consequences of their actions more experientially than
immediate, factual feedback about the student's answer would
allow. The defining feature that makes \stateful{} a game-like
e-learning platform is the fact that it allows dynamical change of the
exercise as whole in response to the student's actions.

The greater power of expression of \stateful{} comes at a cost. For
example, producing the grading logic for a \stateful{} exercise can be
a challenge. It is entirely possible that the student does not visit
some scenes simply because the student is able to correctly solve much
of the exercise at once in an elegant way. Such elegance, however, may
get accidentally punished since such a student would not get points
from the unvisited scenes. Conversely, it is possible that some scenes
get visited many times (possibly with differently randomised parameter
values), but it may not be desirable to give full additional points
for each repetition. For these reasons, the maximum score value of a
\stateful{} exercise cannot plainly be the sum of score values over
all PRTs as is the case in a classical \STACK{} question. We discuss
these aspects in Section~\ref{GradingSec} below.

\begin{figure}[ht!]
\centering
\includegraphics[width=\textwidth]{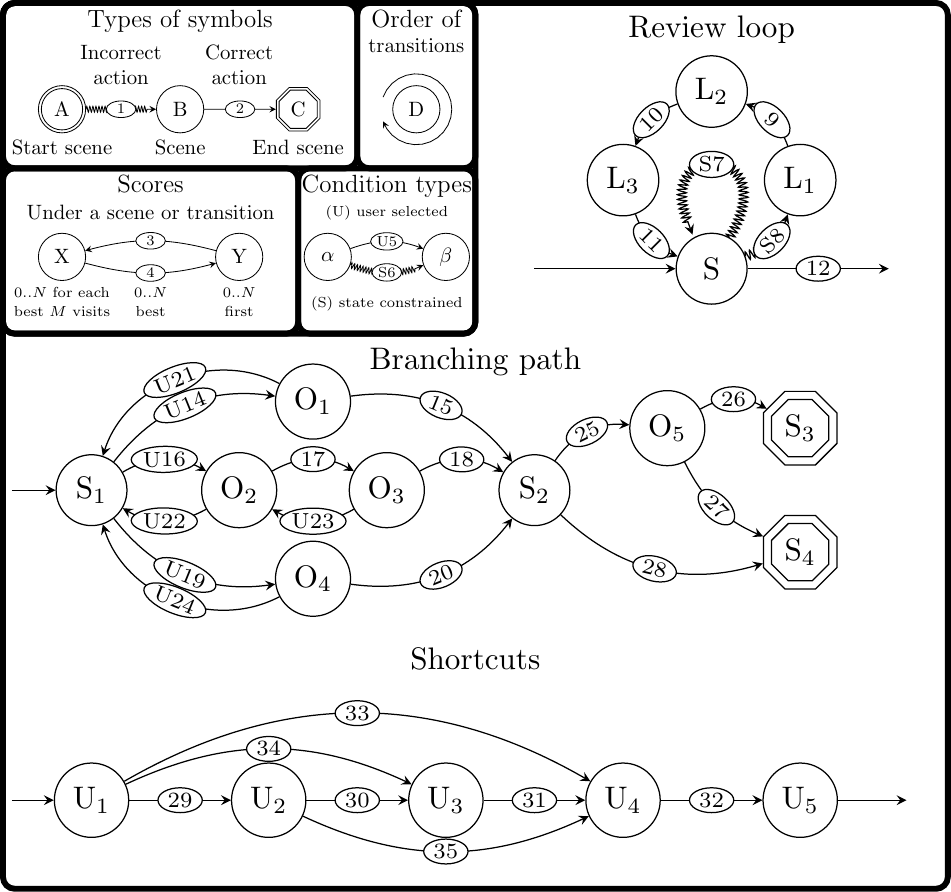}
\caption{Prototypal question shapes for \stateful{}: review loop,
  branching path, and shortcuts. The visual grammar for scene graphs
  is given in the left upper corner.}
\label{shapes}
\end{figure}

The \stateful{} question type has been designed so as to provide
sufficient tools for producing fair grading logics in these
situations.  For example, points can be given for actions within a
scene or for actions that trigger transitions between scenes. The
author explicitly defines how many times one can get the ``same''
points; for example, whether only the first time is counted, or if the
best attempt is rewarded. Alternatively, the author may use additional
state variables to keep track of the student's achievements instead of
giving points at scenes and transitions. In this case, the total score
is constructed during the last transition. This latter approach may be
easier to program, but then the student does not see any motivating
partial scores before reaching the end of the exercise. We point out
that the design of \stateful{} exercise benefits greatly from using
graphical \emph{scene graphs} such as given in Fig.~\ref{shapes} for
three prototypal question shapes to be reviewed below.  Accidents in
grading logic can be detected by describing all sources of points in
such graphical presentations.  Observe that some point sources are
mutually exclusive, e.g., in branching questions as in
Fig.~\ref{shapes}.

Classification of students' behaviour is possible
based on the paths they went through the exercise, or whether they
reached the end scene at all. Additional data for statistical analysis
can be obtained from the actual scene transitions, the changes in the
state within scenes, and the steps on the path where the student
gained points from PRTs. It is worth observing that the expedient
design of a \stateful{} exercise is a prerequisite for later fruitful
path data analysis: The overall sequence of scene transitions should
preferably reflect different problem solving strategies or skill
levels.  It is likely that the students' interaction data contains
surprises to the question author. Hence, the interaction data is
useful for improving the structure and the grading logic of the
exercise. We return to these questions in Section~\ref{MetricSec}.

\subsection{\label{PrototypalSec} Three prototypal question shapes for \stateful{}}

Three prototypal \stateful{} question shapes are illustrated in
Fig.~\ref{shapes}, namely \emph{review loop}, \emph{branching path},
and \emph{shortcuts}. These question shapes make appearances as parts
of more complicated \stateful{} exercises, and we proceed to briefly
discuss them from e-learning material author's point of view.

The first question shape in Fig.~\ref{shapes} is the \emph{review
  loop}.  The question scene S is given in the
beginning. Should the student fail to give an acceptable answer, the
student is required to complete the material in a loop consisting of
multiple scenes (L\textsubscript{1}...).  An acceptable answer is
required in each of these review scenes, and there may be different
loops depending on the nature of student's mistake. Considering the
presentation in Fig.~\ref{shapes}, the transition S7 may be
conditioned so that the student is allowed some number of trials
before being forced to go through the review loop. It may be desirable
that review loop is experienced by the student (at most) once. The review loop 
is the only structure employed by the \SDE{} described in Section \ref{DieselSec}.

The second question shape in Fig.~\ref{shapes} is the \emph{branching
  path}.  Now the students are supposed to complete the task by
letting them choose the problem solving method.  For example, student
could use energy based solution or plainly write the equations of
motion in an elementary theoretical mechanics question.  It is
possible to use multiple branching points in a branching path question
and limit the student's options, if necessary.

Branching paths require some state logic to keep track of the route
chosen by the student. Knowing the route may not be necessary for just
giving the score, but it is indispensable for understanding the
students' typical behaviour patterns. Transitions U14, U16, and U19 in
Fig.~\ref{shapes} represent the students' conscious choice of the
branch they decided to travel. The choice of the branch can also
depend on the features of the students answer as in transitions 25--28
in Fig.~\ref{shapes} which is, in fact, a defining feature of the
question shape \emph{shortcuts} discussed below. Finally, branchings
can be determined based on accumulated data\footnote{Currently, only
  data collected within the same question material can be used. Data
  integration to a separate learning analytics system remains a future
  prospect.} about the student.

The transitions U21--U24 in Fig.~\ref{shapes} represent the student
pressing a "back" button.  Some paths may merge at some point, and the
action of a "back" button is not well defined in such merge points
(unless the student's full route has been stored and made accessible
to the PRT logic by the question author). When the student is allowed
to traverse backwards, it is necessary to keep track of the points
accrued so as to prevent collecting extra points from multiple paths.

The third pattern in Fig.~\ref{shapes} describes \emph{shortcuts}
where transitions 33--35 skip parts of the exercise.  The purpose of
shortcuts is to make the exercise adapt to students' needs by offering
additional feedback and instruction only when it is deemed
necessary. In a typical shortcuts \stateful{} exercise, student is
expected to give the correct answer (such as the value of a definite
integral) but a number of technical steps U\textsubscript{2}, U\textsubscript{3}
(such as integration by parts, change of variables, etc.) are required
to produce it. The pedagogical aim is to learn carrying out these
technical steps, and the correct (partial) answer is used just to
avoid poorly motivated drilling for students that do not benefit from
it. A shortcuts exercise can be designed so that it identifies the
parts of the problem solving process that are difficult for a
particular student, and it then concentrates the training work on
these parts.

\subsection{Visualisation of student data in \stateful{}}

\begin{figure}[ht!]
\centering
\includegraphics[width=\textwidth]{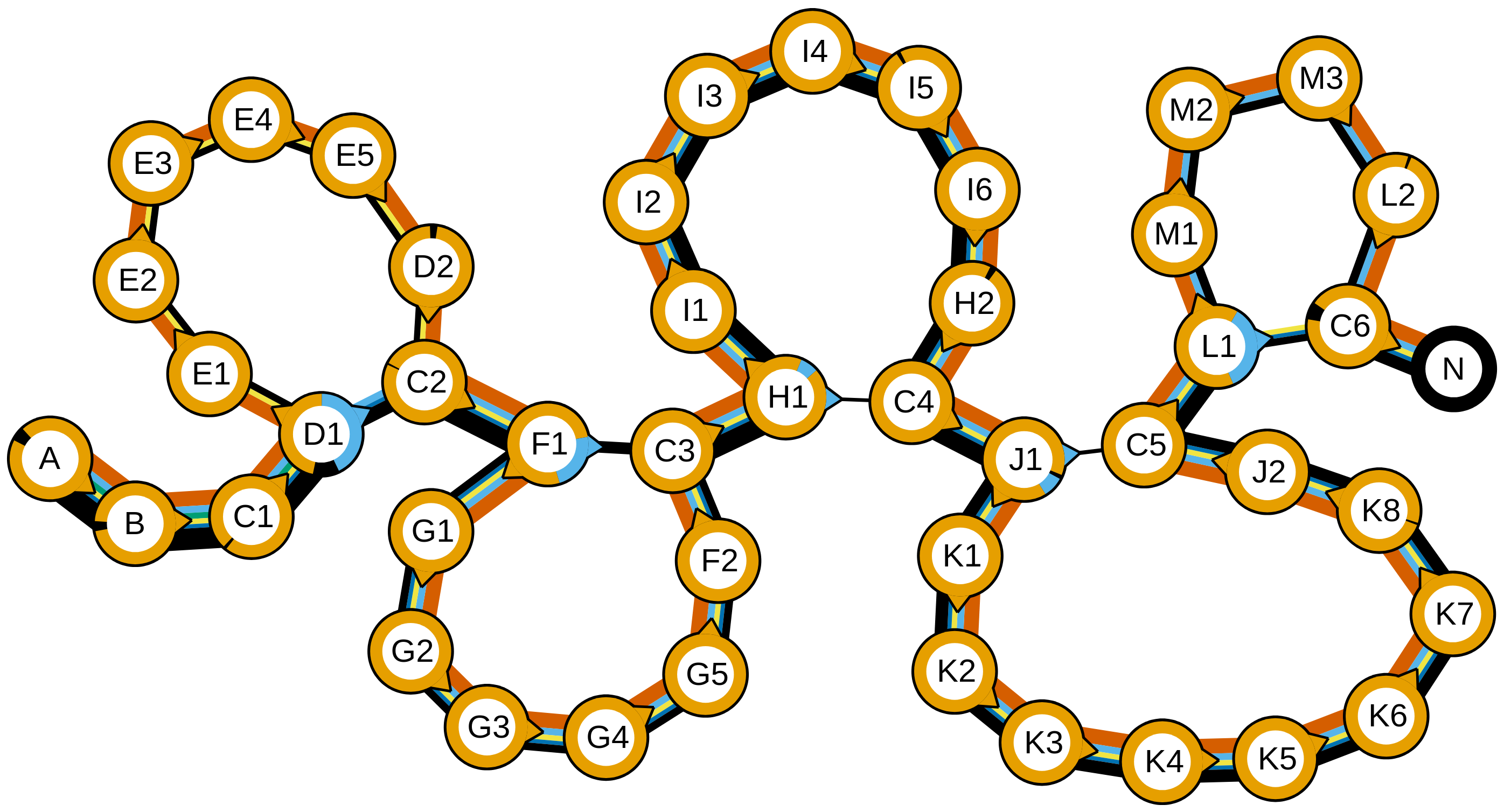}
\caption{\label{path-graph} Student paths through \SDE{} 
consisting of the main path and five review
  loops. Description of the scenes at nodes A, B, C1, D1, etc., are
  given in 
  Section~\protect\ref{ExerciseSec}.  
  The slices around the node indicate the fractions of students
  continuing on specific branches, and the black slice signifies
  attrition. The thickness of the coloured lines between nodes relates
  to the number of students taking the same path through the whole
  question.}
\end{figure}

Understanding the technical function of and the student performance in
a \stateful{} exercise benefits from specific visualisation
tools. Such tools should be applicable to any question shape and all
possible paths that could be taken within the \stateful{} question
type. The flow diagram in Fig.~\ref{path-graph} represents \SDE{} used
in the e-learning experiment of this article, consisting of five
substructures of the review loop type as explained in
Section~\ref{PrototypalSec}. Colours are used to describe the
distribution of the various paths the students took through the
exercise.

Fig.~\ref{path-graph} makes it possible to visually spot the paths
most travelled and the distributions of transitions at the branching
points. The most common paths from the start node A to the end node N
differ only at the first branching point. Student attrition can be
observed since the bundle of paths leaving A is thicker than the
bundle entering N.  Also, branching at nodes D1 and L1 takes place in
a rather balanced way whereas most students at F1, H1, and J1 end up
taking the long route. This may be regarded as an indication of
perceived difficulty of the questions determining the student path at
F1, H1, and J1. These observations may be useful in improving \SDE{}.

The node positioning in Fig.~\ref{path-graph} has been drawn using
Graphviz \cite{Ellson03graphvizand}, and the graph has then been
augmented with custom annotations. The color palette by Bang Wong
\cite{Wong2011} has been used to ensure readability.

\subsection{\label{GradingSec} Grading of \stateful{} exercises}

Grading \stateful{} exercises is more difficult both pedagogically and
technically than grading exams or traditional \STACK{}
questions. Whereas the grading of exams should only measure the
learning outcome in a fair manner, the grading of large \stateful{}
exercises could also pay attention to encouragement of pedagogically
desirable behavioural patterns as well as keeping up the student
motivation to carry on.  Furthermore, it is possible to better detect
and reward diligence, labor, and strategic competence in \stateful{}.
Even then the grading of \stateful{} exercises could contribute to the
final grade of the course as a form of continuous assessment.

The scenes in a \stateful{} exercise are traditional \STACK{} questions
that have their PRTs and  grading logics defined as such. Moreover, state
variables of \stateful{} can be used to realise a wide range of grade
components shared by the scenes. The question author must take
a position in a number of questions in designing a grading logic:
\begin{enumerate}
\item Should alternative but correspondingly parallel parts of the
  exercise have equal point values?
\item Should the review loops have point value? If they should, then
  what proportion of the point value of the direct route?
  Furthermore, if the student collects points from the review loop, is
  they then allowed to gain points from the direct route as well?
\item Should the student be rewarded for activity or success, or to
  some degree for both? Perhaps, the student should be rewarded
  plainly for learning instead of skill.
\item Should wrong attempts (e.g., in MCQs) be penalised, i.e., given
  a negative point value to discourage trial and error?
\item How to prevent the grading logic to be gameable in an
  undesirable manner?
\end{enumerate}
We point out that the question about penalisation applies to all
\STACK{} style grading, and gameability might even be a positively
motivating feature in \stateful{}. Grading and other forms of
rewarding have deep connections to students' motivation and engagement
in the context of gamified learning; see \cite{Alsawaier:2018}.

We proceed to discuss practical aspects of grading on prototypal
question shapes introduced above. One could argue on the basis of
fairness that the student should not be given any extra points for
passing the review loop in Fig.~\ref{shapes}. The student could,
however, obtain points by correctly answering the original question
they was given first time just before entering the review loop.  It is
also possible to set state variables in scenes S7 or S8 of the review
loop in Fig.~\ref{shapes}, and then use them for changing the score at
exiting through transition 12 in order to introduce point value for
working through the review loop.  In general, the alternative paths in
a branching path question in Fig.~\ref{shapes} should be designed in a
reasonably balanced way.  On the other hand, one of the alternative
paths might represent a straightforward, yet a laborious approach
whereas the other path would be elegant but theoretically more
challenging.  It would be difficult to defend why two fully acceptable
solutions should deserve different scores.  One way of justifying such
variance in grading is to allow the student to choose between
different paths after they has been informed about the difference in
difficulty and maximum points.

Unsurprisingly, the grading of a shortcuts exercise has its
challenges, too. A straightforward approach is to subtract from the
maximum score a penalty that is proportional to the amount of extra
instruction and training used. This is rather unmotivating because the
more the student works, the less points the student gets --- making an
effort in studies has value in itself.  Another solution is to allow
the student to make the same shortcuts exercise twice (albeit with
different parameter randomisations), where only the latter attempt is
graded. In all cases, the grading of shortcuts requires striking a
balance between rewarding efforts and skill, and it becomes even more
challenging if the student should be able to follow the development of
their score during the exercise.

We conclude that the students' need for steady motivation to carry on
and complete the task is common to all \stateful{} exercises. The
student should get encouraging continuous feedback, at least, by
getting indication about the level of completion and the points
already accrued. Computer games have engaging scoring logics that the
gamers consider a rewarding part of the gaming experience. Even though
the motivation and the perceptions of fairness are quite different ---
even problematic --- in the e-learning context, some ideas from the
computer game scene may be worthwhile to consider \cite{Alsawaier:2018}.

\section{\label{MaterialSec} Materials}

\subsection{\label{StudentSec} Test subjects}

To obtain accurate information on student activation and
participation, all course participants were used as test subjects, and
they were duly informed about the study. Since this is a deviation
from the principle of informed consent, an ethical review was
conducted prior to starting the research.

Almost all course participants were first-year civil and mechanical
engineering students.  Of the 182 participants, 140 (77\%) were male
and 42 (23\%) female. The median of student age was 20.7 years.  The
student demographics were obtained from the university records, and no
further demographic information was collected.

\subsection{\label{ExerciseSec} Exercise material}

\begin{figure}[ht!]
\centering
  \includegraphics[width=\textwidth]{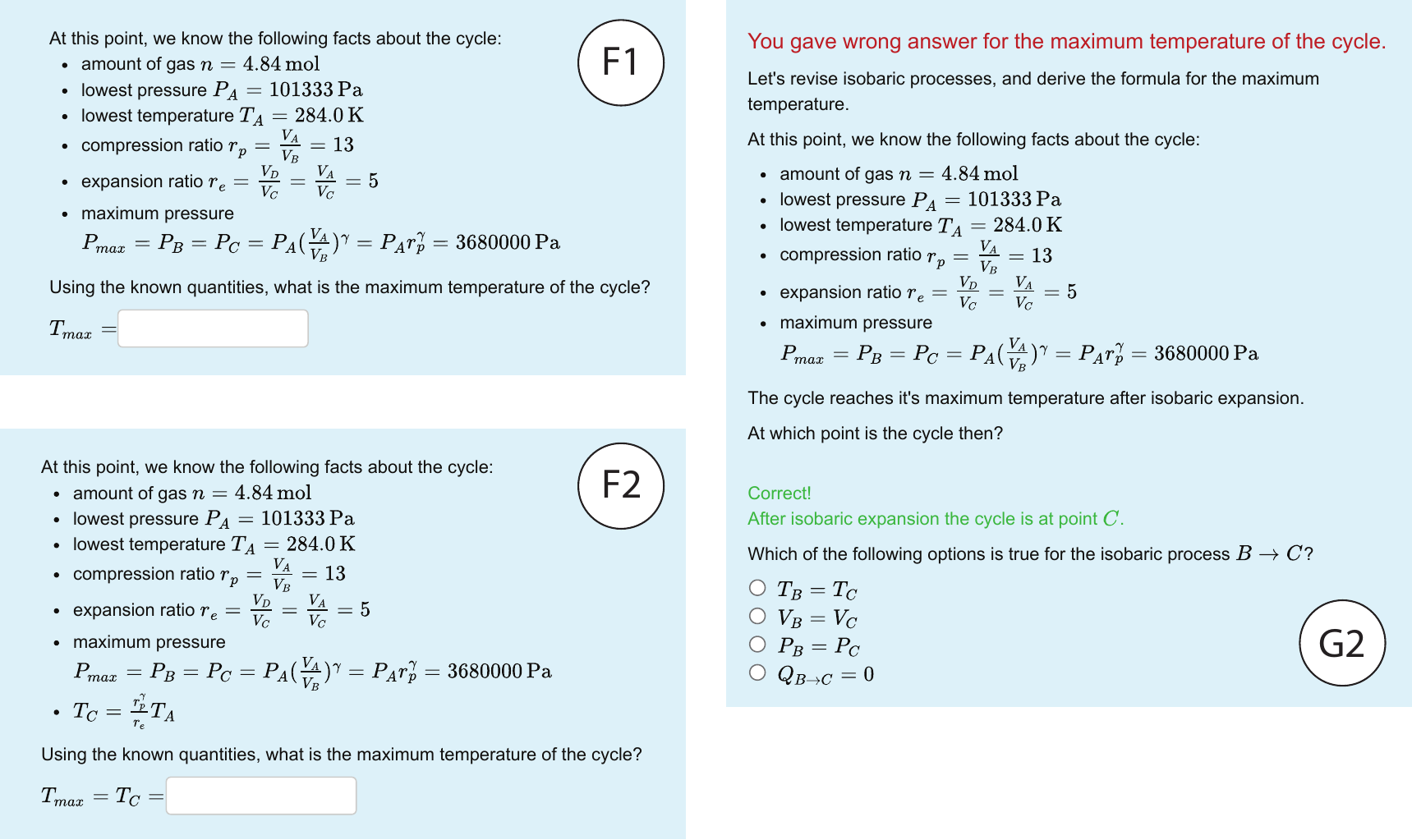}
  \caption{\label{ExMaterial} Three student views from \SDE{}.
    The nodes F1, F2, and G2 are as explained in
    Section~\protect\ref{ExerciseSec}, and they are the same
    as in the visualisation of Fig.~\ref{path-graph}. Node F1
    (top left) is the entry to the scene for solving the
    maximum temperature at the point $C$ of the
    Diesel cycle in Fig.~\ref{Diesel-cycle} (right panel).  In
    case of a wrong answer, the student is directed to the
    review loop G1--G5 in Fig.~\ref{path-graph} consisting of
    several MCQs. Of these, the student view corresponding to
    G2 is shown on the right. After completing the review loop
    G1--G5, the student arrives at node F2 (bottom left) where
    the student tries to solve for the maximum temperature
    again. The formula derived in the review loop G1--G5 for
    the maximum temperature is displayed in scene F2 whereas
    it is not shown in scene F1.}
\end{figure}

The thermodynamical problem of the Diesel cycle was translated into
the \stateful{} format consisting of several scenes as shown in
Fig.~\ref{path-graph}. Only review loops from the prototypal shapes in
Fig.~\ref{shapes} are used in the implementation. The overall
structure of the exercise is linear, but elaborative feedback in the
form of hints help the student forward.  Previous studies have shown
that elaborative feedback is beneficial for learning when administered
after an incorrect answer \cite{attali2017effects}.

The entry scene (A in Fig.~\ref{path-graph}) introduces the Diesel
cycle but has no questions. Then all students are taken to a warm-up
scene B where several MCQs about the relevant thermodynamical
processes are posed. Next, the first main scene C1 is entered where
the problem is set and numerical values are given.  

The first scene where students can end up in different paths is
D1. This is a \STACK{} question where the student is asked to solve for
the maximum pressure of the Diesel cycle.  If the right answer is
given, the student is taken back to a modified main scene C2 from
where the scene F1 is reached. In the scene F1, the maximum
temperature of the cycle is asked by the next \STACK{} question.  In case
of a wrong answer in scene D1, the student is taken to a review loop
E1--E5 consisting of several MCQs on how to calculate the maximum
pressure.  After completing the review loop, the student is asked to
solve again for the maximum pressure in a modified scene D2 with a
\STACK{} question. The difference between scenes D1 and D2 is that in D2
the student has all the results of the review loop E1--E5 at their
disposal.

A similar process is repeated for the remaining four quantities of
interest in the Diesel cycle, i.e., the maximum temperature of the
cycle (F1/F2), heat input into the cycle (H1/H2), heat expelled from
the cycle (J1/J2) and the thermodynamical efficiency of the overall
cycle (L1/L2). All of these have dedicated review loops that trigger
in case of a wrong answer. Finally, the student exits via the final
scene (N).

Some of the student views in \SDE{} are shown in
Fig.~\ref{ExMaterial}.  The nodes F1 and F2 are the views before and
after the review loop for solving the maximum temperature,
respectively. The node G2 is one of the nodes in the corresponding
review loop.  Note that in F2 the student has completed the review
loop G1--G5 and has the information obtained in the loop available.

The question used a simplistic grading logic that gave points from
successfully completing specific key steps, all of these points could
be collected no matter how many attempts it took. All successful
transitions to the C-scenes gave a whole point, starting from 
the completion of the warm-up scene B and ending with the successful
answer to the question about efficiency either directly in
scene L1 or after the review loop in scene L2. In total, a student
could collect six points and is given the cumulative score in real time.

\subsection{Data acquisition from the teaching experiment}

The data collected comes from five major sources: student interaction
with \SDE{}, similar interaction with 27 \STACK{} questions over five
weeks, the manually graded six exam questions, and finally the MPEX
and TCS questionnaires. The exam grading was not repeated with
multiple graders, and possible randomness in it is assumed to be
negligible given the scale of the course.

While most of the 151 students passing the course attempted to do
\SDE{}, only about half of them completed all of
the three questionnaires.  The total number of students who had any
logged activity, in any of the measure contexts, during the course was
182.  Out of these students, 145 started \SDE{},
147 completed the MPEX survey, 107 completed the TCS
pre-questionnaire, and 93 completed the TCS post-questionnaire.  We
analyse the data from all of these activities in a unified manner, and
73 students remain in the intersection suitable for this purpose.

Every input action with \SDE{} described in
Figs.~\ref{path-graph}--\ref{ExMaterial} was logged during the
experiment. In particular, the paths the students took through \SDE{}
as well as the number of failed attempts in specific scenes are a part
of the data set. All data points were time-stamped but the length of
the intervals between consecutive time-stamps need to be interpreted
cautiously, as we do not know whether students were focused on the
task or even present during that time.

Data from \SDE{} consists mainly of two kinds of data items: (i) the
branching points where the student end up in a review loop, and (ii)
the number of extra actions in a given scene. By extra action we mean
any attempts beyond the first input where giving correct answer would
have completed that part of the exercise. The distribution of the
paths through \SDE{} are in Fig.~\ref{path-graph}.

A rather forgiving scoring rules were used in all \STACK{}/\stateful{}
e-learning materials; the students got full points for a correct
answer no matter how many times they tried. While pedagogically justified,
such scoring is far from ideal from a learning analytics point of view as most students will
get full points.  Therefore, we used the numbers of failed attempts to
obtain richer information, e.g., for constructing the fitted metric
discussed below.

\section{\label{MethodsSec} Methods}

\subsection{\label{SOMSec} Clustering of path data}

While each student path through a \stateful{} exercise is unique, they
typically share some common characteristics. Considering \SDE{} in
Section~\ref{ExerciseSec}, we search for underlying behavioural
patterns in students' interactions. For this purpose, the students'
paths were first encoded as vectors consisting of the number of
attempts at each node where interaction is possible. The (almost)
linear design of \SDE{} made such vectorisation quite
straightforward. Resulting \emph{path vectors} can then be clustered
by various methods so as to identify student subgroups and their
behavioural patterns for further analysis.

\begin{figure}[ht!]
\centering
  \includegraphics[width=0.48\textwidth]{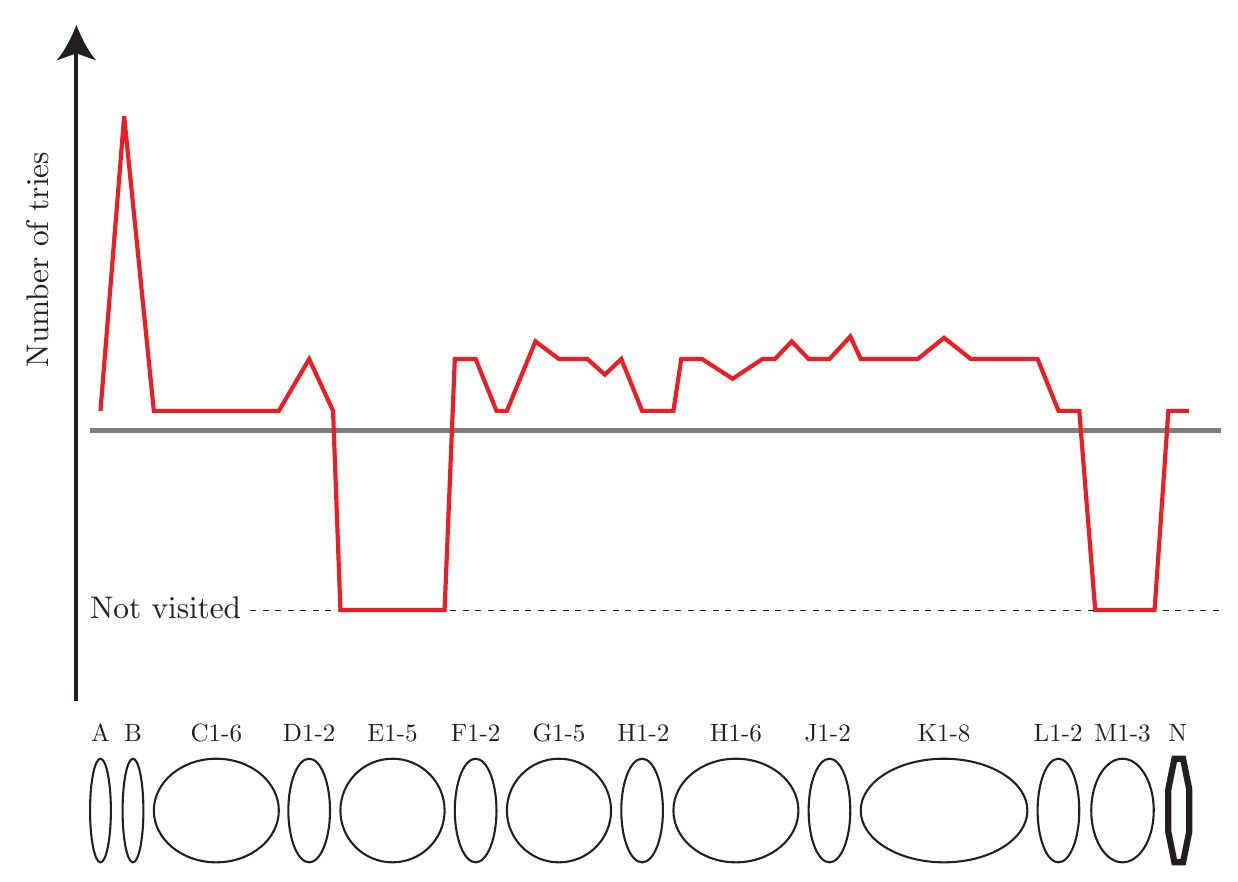}\includegraphics[trim={0 1.6cm 0 0},clip,width=0.5\textwidth]{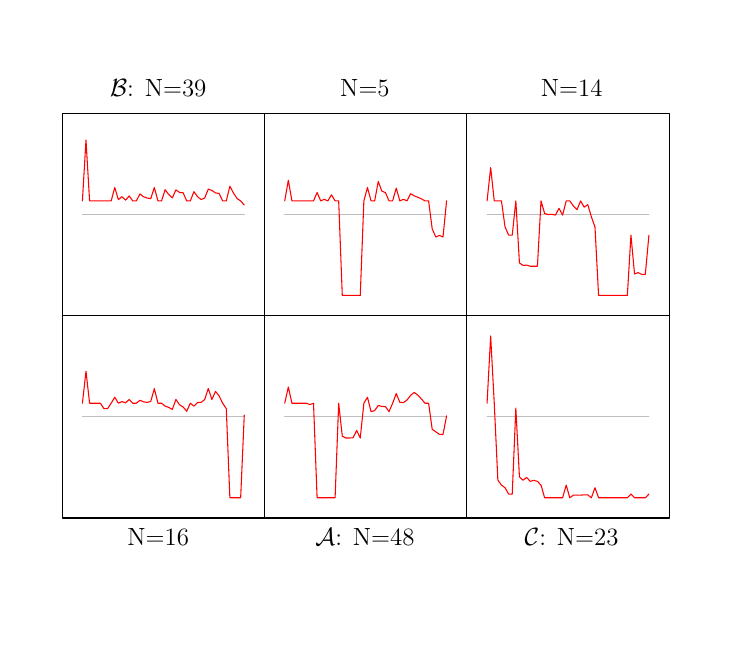}
  \caption{Left panel: An example of a student's path through the
    \SDE{} from the entry scene A to the final scene N, encoded to a
    vector as explained in Section~\protect\ref{SOMSec}.  The temporal
    progression is essentially (but not entirely) linear from left to
    right, and the scene names are as in Fig.~\ref{path-graph}. The
    encoding assumes a negative value in scenes without student's
    interaction. We observe that this student skipped the first and
    the last review loops E and M, respectively.  Right panel: Cluster
    centers of path data vectors of 145 students in six clusters by
    using SOM.  The three largest student groups are denoted by
    \clusterA{}, \clusterB{}, and \clusterC{} as discussed in
    Section~\protect\ref{SOMSec}.
}
\label{codes}
\end{figure}

Self-organising maps \cite{KOHONEN201352} (SOM) were chosen as the
clustering method because it provides a geometric map of the data
where similar vectors tend to be adjacent, helping the interpretation
of observations.  The final size of the adjacency grid (namely 3x2)
was selected as the result of robustness testing where randomly
selected 10\% of the samples were removed from the full set, and the
SOM clustering was repeated for 20 such subsets of the data.  Similar
key clusters were detected in all cases, from which we identified
three major groups of students that were consistently present in all
clustering experiments.  The groups are marked by \clusterA{},
\clusterB{}, and \clusterC{} in Fig.~\ref{codes}. More precisely:

\begin{itemize}
\item[Group \clusterA{}:] Students who skipped the first and/or the
  second review loops, concerning maximum pressure and temperature in
  Diesel cycle, respectively. 48 students.
\item[Group \clusterB{}:] Students who made large numbers of attempts,
  especially in the warmup MCQ part, and continued through all review
  loops. 39 students.
\item[Group \clusterC{}:] Students who gave up without reaching the
  end scene.  23 students.
\end{itemize}
These three student groups \clusterA{}, \clusterB{}, and \clusterC{}
amount to 76\% of the test subject population, 110 students in total.
The three remaining clusters in Fig.~\ref{codes} are of
smaller size, consisting of 35 students that were not readily
classifiable to \clusterA{}, \clusterB{}, and \clusterC{} by SOM.  For
example, there were 16 students who went through all the review loops
except the final loop dealing with thermodynamical efficiency (see
Fig.~\ref{codes}, right panel). We do not consider this cluster as
another student group of interest in this work since its small size
would not allow making statistical inferences.

\subsection{\label{StatSec} Statistical analysis}

The test TCS produces 7 statistical variables, and we only use the
variable \varEmph{MPEXTotal} from MPEX data as explained in
Section~\ref{InstSec}. There is no reason to assume that any of these
variables, nor any other statistical variable such as
\varEmph{ExamTotal}, are normally distributed. Bearing this in mind,
the Kruskal--Wallis test can be used to study whether the mean value
of any such statistical variable is from the same distribution when
conditioned on groups \clusterA{}, \clusterB{}, and \clusterC{}; see
\cite{kruskal1952use}. This process makes it possible to single out
those variables that indicate statistically significant differences
between some of the groups \clusterA{}, \clusterB{}, and
\clusterC{}.

After carrying out the Kruskal--Wallis procedure, the medians of these
component values on the pairs (\clusterA{},\clusterB{}),
(\clusterB{},\clusterC{}), and (\clusterA{},\clusterC{}) can be
compared by pairwise Wilcoxon rank-sum tests, where the significance
value is adjusted with the Bonferroni correction. This pairwise
comparison process is finally extended for the three remaining smaller
clusters to see whether they differ from the groups \clusterA{},
\clusterB{}, and \clusterC{} in terms of MPEX, TCS, or some other
statistical variable available. No reason to join any of the smaller
clusters to any of the groups \clusterA{}, \clusterB{}, or \clusterC{}
emerged from this analysis.

\subsection{\label{MetricSec} Fitted metric}

Compared to the clustering described in Section~\ref{SOMSec}, an
essentially different approach can be taken to make use of the path
and score data from \stateful{}: A parametric model (such as
Eqs.~\eqref{scoringEq}--\eqref{loopscoreEq} below) can be given with
the objective to predict some target, e.g., the outcome of the final
exam. The optimised values of the model parameters can be obtained by
nonlinear optimisation so as to produce a \emph{fitted metric} grading
for the \stateful{} exercise.

Let us first describe one way of defining a score for a \STACK{} style
part, say $P$, of \SDE{} described in
Figs.~\ref{path-graph}--\ref{ExMaterial}.  Given an ordered sequence
of student's attempts, we define the (maximum) score $S \left (P
\right )$ of part $P$ as
\begin{equation} \label{scoringEq}
  S \left (P \right ) 
   = V_P  \cdot \max_i{\left[ \mathrm{v}\left(i\right) \cdot 
     \max{\left(0, 1 - \sum_{j=0}^{i-1} \mathrm{p}\left(j\right)\right)} \right]}
\end{equation}
Here $\mathrm{p}\left(j\right) \in \left [0,1 \right ]$ is the penalty
for the $j$th failed attempt with $\mathrm{p}\left(0\right)= 0$, and
 $\max{\left(0, 1 - \sum_{j=1}^{i-1}
  \mathrm{p}\left(j\right)\right)} \geq 0$ is an expression for the
cumulative penalty factor after $i-1$ failed attempts. The parameter
$V_P$ is the maximum point value of part $P$, and
$\mathrm{v}\left(i\right) \in \left [0,1 \right ]$ gives the fraction
of $V_P$ that the $i$th attempt would have given were it deemed
successful. It is obvious from Eq.~\eqref{scoringEq} what the
student's score in part $P$ would be after getting the acceptable
result after, say, $j = 0,1, \ldots$ unsuccessful attempts.

Considering, for the sake of example, the first review loop consisting
of parts $D_1, E_1, \ldots , D_2$ in Fig.~\ref{path-graph}, we define
its loop score as
\begin{equation} \label{loopscoreEq}
  LC \left ( D \to E \right ) = 
  \begin{cases}
    V_{D_1} & \text{if the answer is acceptable on the first try, and} \\
    V_{D_2} + \sum_{i=1}^4 S \left ( E_i \right )  & \text{if the review loop activates.}
  \end{cases}
\end{equation}
Using the definitions in Eqs.~\eqref{scoringEq}--\eqref{loopscoreEq}
\emph{mutatis mutandis}, one can define scoring function for the whole
\SDE{} described in Section~\ref{ExerciseSec} for the Diesel
cycle. This plainly amounts to summing over all review loops and the
remaining scenes in Fig.~\ref{path-graph}. We emphasise that the
scoring function could be defined in many other ways different from
Eqs.~\eqref{scoringEq}--\eqref{loopscoreEq}, and additional
constraints can be imposed such as never getting a better score from a
review loop than from solving the problem directly.

We now proceed to define the particular type of fitted metric that is
associated to the scoring function defined above. To reduce the number
of optimisation parameters, we simplify Eq.~\eqref{scoringEq} by
allowing the penalty only have constant values
$\mathrm{p}\left(j\right) = p_P$ for each part $P$. Hence, by
Eq.~\eqref{scoringEq}, the score $S \left (P \right )$ becomes a
function of only two parameters $p_P$, $V_P$ and the function
$\mathrm{v}$. Similarly, the total score of the
\stateful{} exercise will be a function of the parameter families $\{
p_{P} \}_P$, $\{ V_{P} \}_P$, and $\{\mathrm{v} \left(j\right ) \}_j$
that are used for nonlinear optimisation.

We point out that the purpose of the fitted metric is not to grade
students in an obscure manner but to extract useful information from
the \stateful{} data in relation to the actual grading of the
course. Indeed, by examining the optimal parameter values thus
obtained, conclusions can be made about which parts of and behaviours
in the \stateful{} exercise best predict students' success in the
final exam.  Such understanding may have value in improving the
e-learning material to better achieve the learning objectives.

\section{\label{ResultsSec} Results and  discussion}

\subsection{\label{ClusterSec} Statistical analysis of  student  groups}

\begin{figure}
  \centering
    \includegraphics[width=0.9\textwidth]{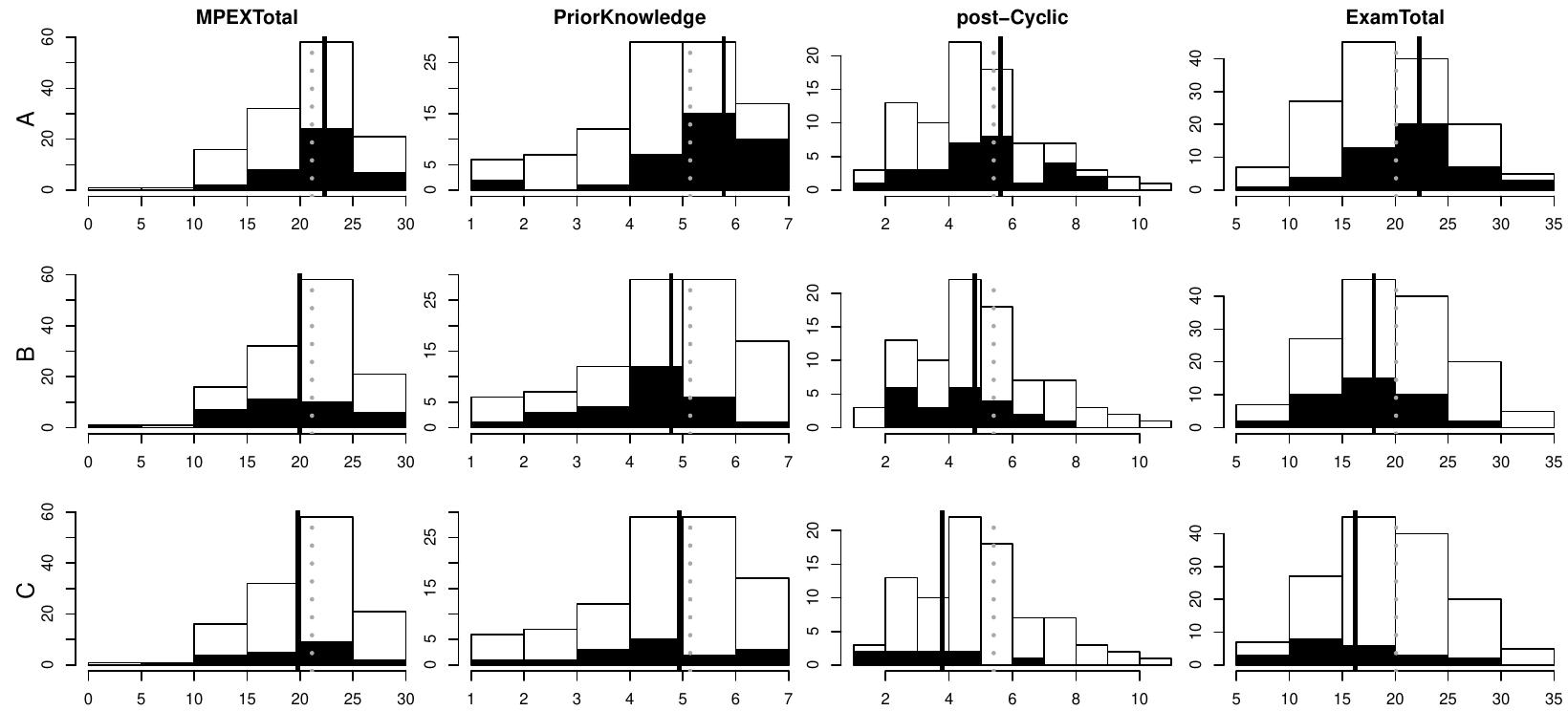}
    \caption{\label{clusters-1} Student numbers in each student group
      \clusterA{}, \clusterB{}, and \clusterC{} (from top to bottom)
      as a function of the statistical variables \varEmph{MPEXTotal},
      \varEmph{PriorKnowledge}, \varEmph{post-Cyclic}, and
      \varEmph{ExamTotal} (from left to right).  In each histogram,
      the white bars show the distribution of entire population and
      the black bars describe the student group in question. The
      dotted vertical line is the average of the statistical variable
      in the entire population, and the continuous vertical line is
      the average of the statistical variable in the student group.}
\end{figure}

\begin{figure}[h!]
  \centering
    \includegraphics[width=\textwidth]{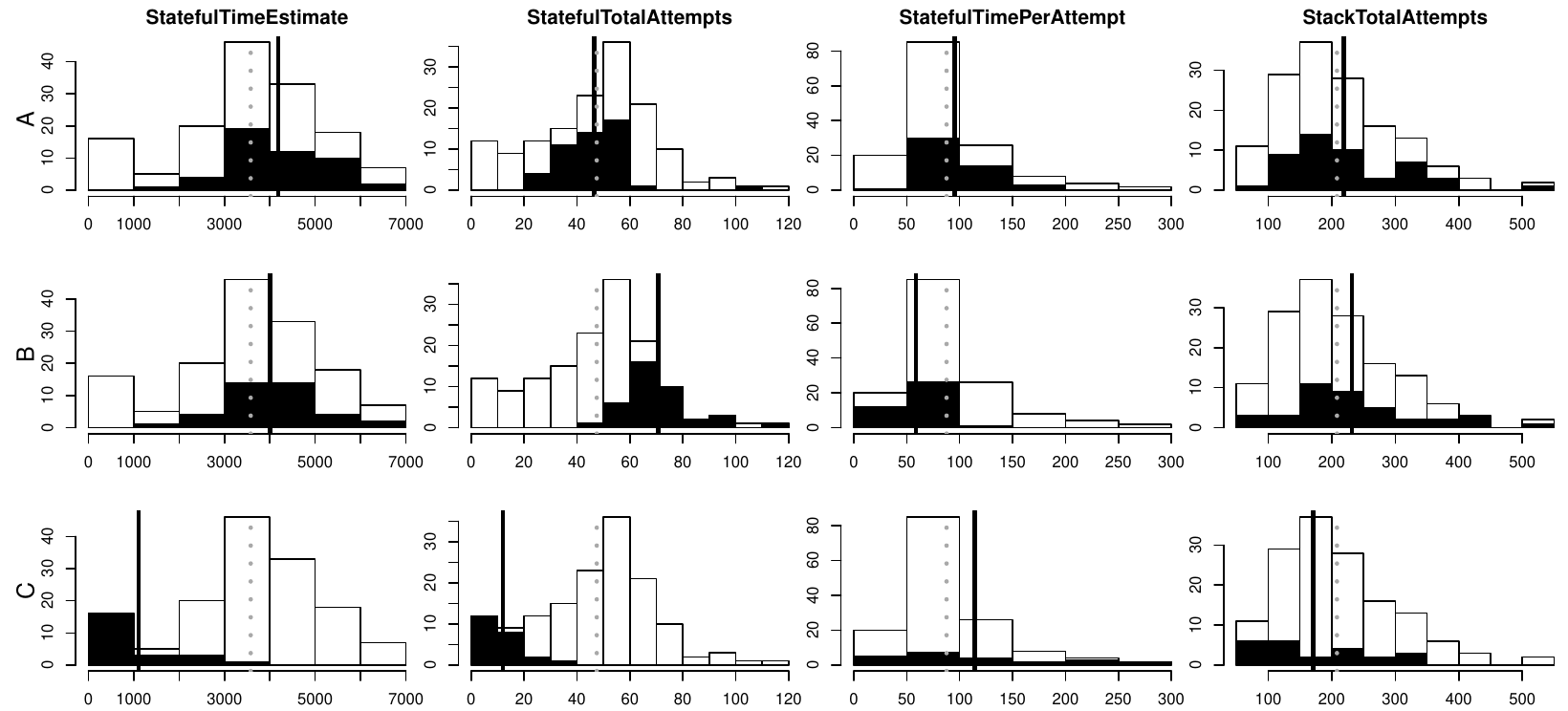}
    \caption{\label{clusters-2} Student numbers
      within each student group \clusterA{}, \clusterB{}, and
      \clusterC{} (from top to bottom) as a function of the
      statistical variables \varEmph{StatefulTimeEstimate},
      \varEmph{StatefulTotalAttempts},
      \varEmph{StatefulTimePerAttempt}, and
      \varEmph{StackTotalAttempts} (from left to right). The unit of
      time is seconds, the average completion time being about one hour. 
      The presentation is similar as in Fig.~\protect\ref{clusters-1}.
}
\end{figure}

The clustering experiment of Section~\ref{SOMSec} indicates that a
large majority of the students can be robustly classified into one of
the three groups \clusterA{}, \clusterB{}, and \clusterC{} based on
their interaction data with \SDE{}. In
Fig.~\ref{clusters-1} statistics involving MPEX and TCS test results
are given, whereas behavioural statistics related to \stateful{} and
\STACK{} are shown in Fig.~\ref{clusters-2}.

We consider next whether these groups are different based on the MPEX
variable \varEmph{MPEXTotal} (evaluating attitudes towards learning
physics) and TCS (assessing conceptual competence in thermodynamics)
reviewed in Section~\ref{InstSec}.\footnote{Recall that TCS was
  carried out twice, before and after \SDE{}.} Another interesting
task is to consider to what extent groups \clusterA{}, \clusterB{},
and \clusterC{} differ from each other in terms of MPEX and TCS.
Concerning the attitudes towards physics as measured by
\varEmph{MPEXTotal}, the three groups \clusterA{}, \clusterB{}, and
\clusterC{} have somewhat different scores, with \clusterA{}
indicating more positive attitudes than \clusterB{} and \clusterC{};
see Fig.~\ref{clusters-1} (leftmost column). According to the
Kruskal--Wallis test, this observation is not statistically significant
at the $p = 0.05$ level ($p= 0.093$).  Hence, we cannot state whether
the students in \clusterA{} have more expert-like beliefs than their
peers in \clusterB{} and \clusterC{} based on this data set.

We observe similar differences between groups in conceptual knowledge
as measured by TCS. As shown in Fig.~\ref{clusters-1} (second column
from left), students in group \clusterA{} show the highest average
score in the \varEmph{PriorKnowledge}. Students in groups \clusterB{}
and \clusterC{} do not differ much in \varEmph{PriorKnowledge} but the
group \clusterC{} has a little higher average score of these two.  A
Kruskal--Wallis test confirms that there are signicant differences in
\varEmph{PriorKnowledge} scores between these three groups
($p=0.0014$).  Further pairwise Wilcoxon tests confirm that the
difference between groups \clusterA{} and \clusterB{} is statistically
significant ($p=0.001$) but not between groups \clusterA{} and
\clusterC{} at the $p = 0.05$ level ($p=0.11$). There is no
statistical difference in \varEmph{PriorKnowledge} (nor any other
statistical variable given in Fig.~\ref{clusters-1}) between groups
\clusterB{} and \clusterC{} ($p=1.00$).
Thus, the students in group \clusterA{} begin their studies with a higher level of knowledge from the upper secondary school.

Two other tests carried out before \SDE{},
namely \varEmph{pre-Adiabatic} (questions 13-19) and
\varEmph{pre-Cyclic} (questions 20-31), do not show differences
between groups \clusterA{}, \clusterB{} and \clusterC{} in score or in
statistics. The \varEmph{pre-Relevant} (questions 8-35) test shows
slightly lower score for the students in group \clusterB{} than groups
\clusterA{} and \clusterC{}, but the difference is not statistically
significant. We conclude that the level of knowledge of thermodynamics
is similar in all groups before instruction in the course.

The post-TCS test results, on the contrary, indicate different
behaviour.  In the \varEmph{post-Relevant} (questions 8-35) test,
group \clusterA{} has the highest average score, and the score for
students in group \clusterC{} is the lowest. However, the number of
students who completed the post-TCS questionnaire is rather small,
especially in group \clusterC{}. The \varEmph{post-Adiabatic}
(questions 13-19) test gives the same statistical characteristics for
groups \clusterA{} and \clusterB{}, the score of group \clusterC{}
being considerably lower. The most interesting case is the
\varEmph{post-Cyclic} (questions 20-31) test in which the best average
score is for the students in group \clusterA{}, and the lowest score
is in group \clusterC{}; see Fig.~\ref{clusters-1}
(second column from right).  Furthermore, Kruskal--Wallis test confirms
that there are statistically significant differences in the
\varEmph{post-Cyclic} test results between the groups
($p=0.027$). Further, pairwise Wilcoxon tests confirm that the
difference between groups \clusterA{} and \clusterC{} is statistically
significant ($p=0.044$) but not between groups \clusterA{} and
\clusterB{} ($p=0.34$). The students in groups \clusterA{} and
\clusterB{} went through \SDE{} completely, and they thus learnt more
about the cyclic processes than the students in group \clusterC{} who
did not finish the exercise at all.  The \varEmph{pre-Cyclic} test
carried out before \SDE{} showed that all the three groups
\clusterA{}, \clusterB{}, and \clusterC{} have similar statistical
characteristics, whereas the \varEmph{post-Cyclic} test shows a
completely different picture as can be seen in Fig.~\ref{clusters-1}. 
We conclude that the clustering of
\stateful{} path data is able to detect student groups whose response
to instruction on thermodynamical cyclic processes is essentially
different.

Regarding the way how students interact with \SDE{} and \STACK{}, we
observe first that groups \clusterA{}, \clusterB{}, and \clusterC{}
differ from each other significantly using the Kruskal--Wallis test
based on all the four statistical variables shown in
Fig.~\ref{clusters-2} (with $p < 0.033$ for all variables). This
supports the division of the student population in groups by SOM.  As
can be seen in Fig.~\ref{clusters-2} (leftmost column), the students
who spent longest time on the \stateful{} question are those in group
\clusterA{}, although the difference to group \clusterB{} is small and
not statistically significant. Group \clusterC{} spent shortest time
in \SDE{}; a consequence of the fact that only one student in
\clusterC{} finished it.  The most successful group \clusterA{} had
the highest and group \clusterC{} the lowest average exam score (see
Fig.~\ref{clusters-1}, rightmost column).  The difference in the exam
score in Fig.~\ref{clusters-1} is statistically significant
(Kruskal--Wallis $p<0.001$) between group \clusterA{} and the other
groups (Wilcoxon $p<0.001$ for group \clusterB{} and $p<0.001$ for
group \clusterC{}), but not between groups \clusterB{} and \clusterC{}
(Wilcoxon $p=0.46$).

\begin{figure}[h!]
  \centering
    \includegraphics[width=\textwidth]{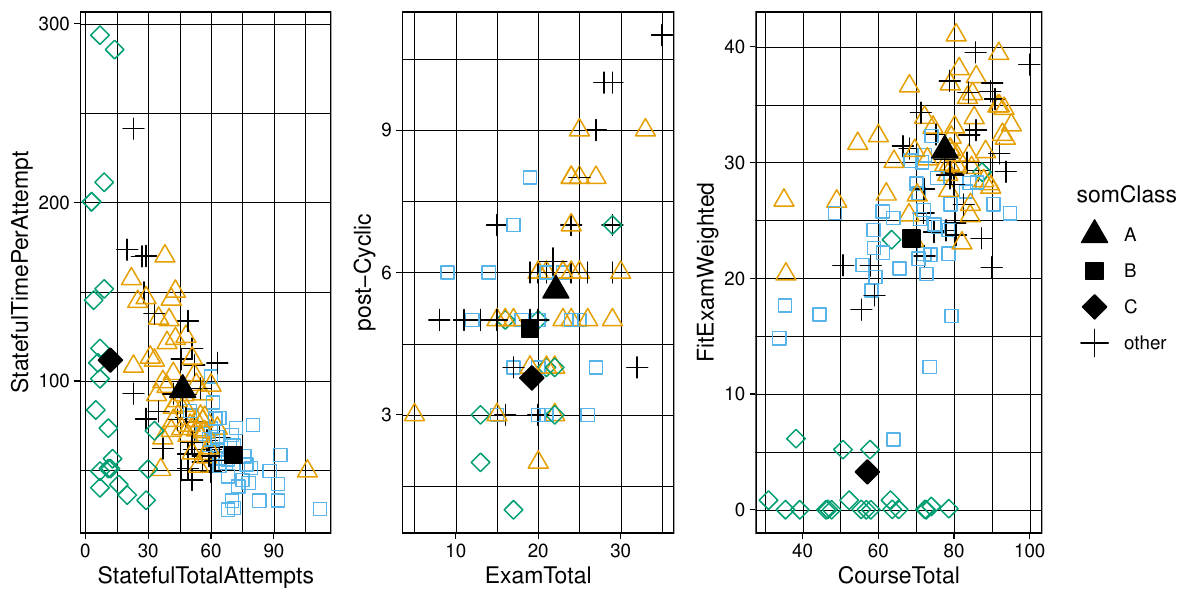}
    \caption{\label{correlations-1} Scattergrams of variable pairs
      where students' behaviour in \stateful{} has been described by
      groups \clusterA{}, \clusterB{}, and \clusterC{}.  The group
      centroids have drawn with filled symbols.  Left panel: Different
      strategies of time usage illustrated by
      \varEmph{StatefulTimePerAttempt} versus
      \varEmph{StatefulTotalAttempts}.  Middle panel: A good learning
      outcome in cyclic processes in TCS is associated to success in the
      final exam as shown by \varEmph{post-Cyclic} versus
      \varEmph{ExamTotal}.  Right panel: The optimal fitted metric
      predictor \varEmph{FitExamWeighted} is plotted versus
      \varEmph{CourseTotal}. The points of the final exam (with an
      extra weighting) have been used as an optimisation target in the
      fitted metric model described in Section~\ref{MetricSec}.}
\end{figure}

We proceed to analyse further the observed differences between the
student groups \clusterA{}, \clusterB{}, and \clusterC{} in terms of
correlation analysis.  We consider next the time usage and the total
number of attempts in \SDE{}.  The number of
\varEmph{StatefulTotalAttempts} for group \clusterB{} is the highest,
for group \clusterA{} considerably lower, and for group \clusterC{}
clearly the lowest; see Fig.~\ref{clusters-2} (second column from
left). The statistical variable \varEmph{StatefulTimePerAttempt} in
Fig.~\ref{clusters-2} (second column from right) is the average time a
student used in each attempt, and it is plainly
\varEmph{StatefulTimeEstimate} divided by
\varEmph{StatefulTotalAttempts}.  The statistical variables
\varEmph{StatefulTotalAttempts} and \varEmph{StatefulTimePerAttempt}
from \SDE{} have a clear negative correlation
(Pearson $r = -0.56$); i.e., the more attempts the student makes, the
less time is used for each attempt; see Fig.~\ref{correlations-1}
(left panel). Such a rough inverse relation is, of course, to be
expected since the students have their time constraints. However, the
typical behaviour in groups \clusterA{}, \clusterB{}, and \clusterC{}
is quite different as can be seen in Fig.~\ref{clusters-2}. The
highest value for \varEmph{StatefulTimePerAttempt} is for group
\clusterC{}, and the value for group \clusterA{} is only slightly
lower. The values of \varEmph{StatefulTimePerAttempt} for group
\clusterB{} is almost half of the values of the other two groups.  The
hard working students in group \clusterB{} seem to make more attempts
than students in group \clusterA{}, and they continue until they find
the correct answer.  These differences can also be seen in the
placements of groups \clusterA{}, \clusterB{}, and \clusterC{} in
Fig.~\ref{correlations-1} (left panel). The raw data from \stateful{}
shows that students in group \clusterB{} go through all the loops in
\SDE{} (see Fig.~\ref{codes}), but the
success in the final exam (see Fig.~\ref{clusters-1}) remains
considerably weaker compared to the students in group \clusterA{}.

The nodes of \SDE{} are quite typical \STACK{} questions, and it is
hence not surprising that groups \clusterA{}, \clusterB{}, and
\clusterC{} display similar behaviour in also the traditional \STACK{}
exercises used in the course.  Indeed, group \clusterB{} has the
highest average value in \varEmph{StackTotalAttempts} and \clusterA{}
has a somewhat lower values, whereas the value for group \clusterC{}
are clearly the lowest; see Fig.~\ref{clusters-2} (rightmost
column). None of the differences between the groups in
Fig.~\ref{clusters-2} (rightmost column) are statistically significant
by pairwise Wilcoxon tests, but there are similar (albeit weaker)
behavioural differences between the groups as was observed in
Fig.~\ref{clusters-2} (second column from left) for \stateful{}. It is
worth observing that \varEmph{StackTotalAttempts} on groups
\clusterA{} and \clusterB{} are practically uncorrelated, reflecting
the behavioural differences elicited by interaction with
\STACK{}/\stateful{} materials.  We conclude that the weak commitment
in group \clusterC{} to e-learning is not limited to the novel \SDE{},
but the underlying reasons remain unknown in this study.

There is a reasonably good correlation (Pearson $r = 0.50$) between
the scores of the final exam and the \varEmph{post-Cyclic} TCS
questions, indicating that the students who have learnt well cyclic
thermodynamical processes in the course (including \SDE{}) have also
succeeded in the final exam; see Fig.~\ref{correlations-1} (middle
panel).  We point out that the students in group \clusterA{} are
mainly located in the upper-right quarter while the students in group
\clusterC{} are mainly in the lower-left quarter in
Fig.~\ref{correlations-1} (middle panel).

As already observed in Fig.~\ref{clusters-1} and its statistical
analysis, the \varEmph{pre-Cyclic} TCS questions do not indicate
statistical differences between the student groups \clusterA{},
\clusterB{}, and \clusterC{} whereas consistent differences appear
both in \varEmph{post-Cyclic} TCS questions and in the final exam.  It
seems plausible that \SDE{} has had a positive role in the learning of
the cyclic processes, but the additional learning benefit of using
\SDE{} cannot be quantified since a controlled comparison group
(without \stateful{} exposure) was not used. We conclude that
learning, indeed, has taken place, and that the different behaviour
patterns observed in \SDE{} are associated to different learning
outcomes in a robust and consistent manner.

\subsection{Observations about the fitted metric}

As pointed out in Section~\ref{StatefulSec}, grading \stateful{}
exercises can be quite a challenge. We introduced one possible
parametric grading model for \stateful{} exercises with review loops
given by Eqs.~\eqref{scoringEq}--\eqref{loopscoreEq} in
Section~\ref{MetricSec}.  If other parts of the course (such as the
traditional exercises, laboratory work, or the final exam) already
have numerical grading, the model coefficients in
Eq.~\eqref{scoringEq} can be determined so that the resulting grading
of the \stateful{} exercise optimally reflects the success of the
student in such other parts. We call such model-based grading produced
by, e.g., Eqs.~\eqref{scoringEq}--\eqref{loopscoreEq} with optimised
coefficients \emph{fitted metric}. We proceed to study whether the
fitted metric, indeed, correlates with the more traditional grading,
and whether it is able to separate the student groups \clusterA{},
\clusterB{}, and \clusterC{} from each other.

Using nonlinear optimization in Octave, the 72 model parameters in
Eqs.~\eqref{scoringEq}--\eqref{loopscoreEq} (when applied to \SDE{})
were optimised so as to best predict the somewhat modified final exam
grading where the weight of the problem involving cyclic processes was
doubled. The outcome of the fitted metric is the statistical variable
\varEmph{FitExamWeighted}, and the scattergram of
\varEmph{CourseTotal} and \varEmph{FitExamWeighted} is given in
Fig.~\ref{correlations-1} (right panel). We observe first that these
statistical variables have a strong correlation (Pearson $r = 0.61$)
over the whole student population.  Secondly, it can be clearly
seen that most of the
students in group \clusterC{} did not finish \SDE{}: the fitted metric fails to predict \varEmph{CourseTotal} for
these students because the information in \stateful{} learning data is
simply not complete enough.  Thirdly, the \stateful{} learning data
contains a fair amount of useful information about students in groups
\clusterA{} and \clusterB{} in contrast to group \clusterC{} where the
fitted metric fails to have skill of prediction. However, the fitted
metric successfully separates groups \clusterA{}, \clusterB{}, and
\clusterC{} in Fig.~\ref{correlations-1} (right panel). It is also
worth observing that the students not classified to any of the groups
\clusterA{}, \clusterB{}, and \clusterC{} appear uniformly among
\clusterA{} and \clusterB{} but not at all among \clusterC{} in
Fig.~\ref{correlations-1} (right panel) whereas the same students
appear among \clusterA{} but not among \clusterB{} and \clusterC{} in
Fig.~\ref{correlations-1} (left panel).

Similar fitted metric analysis can be carried out by using, e.g., the
total score instead of the (modified) grade of the final exam as the
optimisation target for
Eqs.~\eqref{scoringEq}--\eqref{loopscoreEq}. Changing the target for
parameter computations does not seem to essentially change the
observations, supporting the claim that groups \clusterA{},
\clusterB{}, and \clusterC{} --- produced by the entirely different
method SOM --- reflect genuine learning related differences in the
student population.

We discuss next how the fitted metric can be used to compare the
quality and the relative significance of the parts of \SDE{}. The
model parameter optimisation described above leads to a grading that
is mathematically aligned with the (implicit or explicit) principles
used for grading the final exam.  \SDE{} consists of six parts listed
in Table~\ref{fitted} below, of which five parts have review loops. Of
these parts, the second and third (i.e., $P_{\max}$ and $T_{\max}$)
get much higher percentage of the total points in fitted metric
grading than any other parts, indicating that the student activity
there is particularly well-aligned with the learning objectives as
measured in the final exam.

With the exception of the first and the last parts with review loops
(i.e, $P_{max}$ and \emph{Efficiency}), the (maximum) point scores are
the same for both the direct route and the review loop. This can be
seen as success in \SDE{} design since penalising necessary extra
work carried out in review loops may not be desirable.  This is in
contrast to the last part \emph{Efficiency} where ending up in the
review loop leads to low maximum points in fitted metric grading
(i.e., direct route $16.1 \%$ vs. review loop $4.3 \%$). Such a
drastic difference, however, need not be due to poor design of the
review loop. It is thinkable, for example, that the review loop in
\emph{Efficiency} does not strongly relate to what actually gets
measured in the final exam. Another possible explanation is that
students' concentration at the end of \SDE{} has already lapsed,
and the activity data from the last review loop may reflect poorly
their best performance. It can be observed from Table~\ref{fitted}
that the maximum points have a general decreasing tendency towards the
end of \SDE{} which may indicate increasing randomness in
students' reactions due to exhaustion.

\begin{table}[H]
\caption{\label{fitted} Point value distribution for the \SDE{},
  determined from the grading by a fitted metric whose optimisation
  target is a weighted sum of the final exam question scores.  The
  percentage given indicates the fraction of the maximum total point
  value of the part in question. The last column refers to scenes in
  Fig.~\ref{path-graph}.}
\begin{center}
\begin{tabular*}{0.67\linewidth}{rrrr}
  \hline
Exer. part & Direct route & Review loop (max.) & Relevant scenes \\ 
 \hline
  Intro & \(4.6\%\) & NA/same & B \\
  \(P_{max}\) & \(24.2\%\) & \(21.5\%\) & D1, or E1--E4 and D2 \\ 
  \(T_{max}\) & \(22.2\%\) & \(22.2\%\) & F1, or G1--G4 and F2 \\ 
  \(Q_{in}\) &  \(16.8\%\) & \(16.8\%\) & H1, or I1--I5 and H2 \\ 
  \(Q_{out}\) & \(16.0\%\) & \(16.0\%\) & J1, or K1--K7 and J2 \\ 
  Efficiency &  \(16.1\%\) & \(4.3\%\) & L1, or M1--M2 and L2 \\ 
  \hline
  Total & \(100\%\) & \(85.5\%\) \\ 
   \hline
\end{tabular*}
\end{center}
\end{table}

\section{\label{DiscConcSec} Conclusion}

We have introduced a novel method for teaching thermodynamics of
cyclic processes in elementary university physics.  The method is
based on an interactive, adaptive e-learning material on Diesel cycle
developed on the \stateful{} platform: namely the \stateful{} Diesel
exercise (\SDE{}). A teaching experiment was carried out using 182
first-year students of mechanical engineering as test subjects. A
large amount of data was collected from the behaviour of students in
\SDE{} and their performance in other parts of the course.

It was observed that the learning data from \SDE{} is rich enough for
robustly classifying 76\% of the students into three groups
\clusterA{}, \clusterB{}, and \clusterC{} by using Self-Organising
Maps (SOM) as a clustering algorithm. These groups have statistically
significant differences both in other course metrics and also in
observed behaviour in problem solving situations provided by
\stateful{}.  There are even some differences in students' general
attitudes towards learning physics between the three groups but a
statistically significant result cannot be given about it.

Students in group \clusterA{} navigate through the exercise avoiding
some of the review loops and use only a moderate amount of attempts
even though extra attempts are not penalised in the exercise.  Group
\clusterA{} is overall very successful in the course, and it is not a
surprise that they also do well in \SDE{}.  Students in group
\clusterB{} clearly rely on solving the exercise using a lot of
trials. Lack of penalties in review loops certainly encourages this
behaviour but as the group \clusterA{} shows, this is not the way all students respond to \SDE{}. The expert-like attitudes (as measured by MPEX) in
group \clusterB{} are not statistically different from those in group
\clusterA{}. However, the overall prior knowledge (as measured by TCS)
is lower in group \clusterB{} than in group \clusterA{} but these
groups do not show statistically significant differences about the new
concepts relevant to \SDE{}. The students in group \clusterB{} also
tend to have lower performance in the final exam. An interesting
future experiment would be to use penalties in a \stateful{} exercise
and observe if the behaviour in or the size of group \clusterB{}
changes.

The third group \clusterC{} is most problematic from teacher's point
of view. Their engagement in exercises, including \SDE{}, is lower
than for the two other groups. This is reflected in lower scores in
all parts of the course, including the final exam. Moreover, there is
less data available for group \clusterC{} than for groups \clusterA{}
and \clusterB{}, making the statistical analysis less
conclusive. Engaging students in group \clusterC{} to take part in
course activities remains a challenge.

In addition to the clustering approach by SOM, an entirely different
(yet consistent) kind of view into \stateful{} learning data was
acquired by fitted metric analysis. There, a parametric model was
proposed so as to optimally predict some target statistical variable
that is independent of \SDE{}. In this work, we used a weighted
version of the final exam points as the target for numerical model
parameter estimation. The model prediction of, e.g., the total points
(determining the student's final grade) is surprisingly good even
though the predictive skill of the fitted metric fails in the
challenging group \clusterC{}. This outcome is particularly remarkable
because such predictive skill is based on multiple facets of a
singular subject matter (i.e., the Diesel cycle) measured over a
singular work unit (i.e., \SDE{}) which is a just small part of the
course. It is, therefore, thinkable that the \stateful{} learning data
contains a glimpse into general learning and problem solving skills
and thus provides a prediction about the future success of the
student. It remains open how these observations would change if
several \stateful{} exercises of various design would be used, if a
more complicated fitted metric model than
Eqs.~\eqref{scoringEq}--\eqref{loopscoreEq} were developed, or if the
learning analytics were aided by some modern type of machine learning
model on a much larger student population.

The fitted metric can be used as grading for \stateful{} exercises
with some caution. Even though fitted metric is mathematically aligned
with its target, the grading by it may fail to be comprehensible or
fair to the student in the context of the exercise. Moreover, the
fitted metric may have adapted to, e.g., the effects of students'
exhaustion towards the end of the exercise which should not affect
unbiased grading. The fitted metric, however, can be a starting point
of a usable grading if the materials author manually modifies what
needs to be changed.

Upper secondary school and university education have been in a state
of constant reform during the last ten years in Finland. At the same
time, a remarkable cultural change has taken place where computers and
mobile devices (including ubiquitous smart phones) have become
essential elements in many aspects of education and everyday life. The
traditional classroom and lecture hall seem to have lost some of their earlier
appeal to students. In addition, university education is now given to
a larger part of the age group while the age groups are getting
smaller due to demographic transition. Keeping up the level of
learning outcome, the quality of education, and student motivation in
STEM fields would greatly benefit from fresh new ideas since the
usable resources for teaching are unlikely to increase. Considering
this background, technologies such as STACK and \stateful{}, combined
with learning analytics by modern data science tools, seem one
promising way to proceed. Such platforms may provide students an
attractive user experience with interactive, even game-like e-learning
materials.  Access to a deeper learning data helps in developing more
compelling e-learning materials and detecting weak spots in course
design. Finally, digital platforms are expected to free teachers'
working time that can be used more fruitfully in individual contact
teaching and tutoring for those who require it, instead of, e.g.,
manually marking solutions from exercise sessions.

\section*{Acknowledgements}
We wish to thank Dr.~P.~Alestalo and Dr.~R.~Kangaslampi for valuable
discussions and A.~Al-Adulrazzaq for help with the MPEX translation.

\bibliographystyle{cas-model2-names}

\bibliography{publications}

\end{document}